\documentclass[aps,prb,twocolumn,showpacs,superscriptaddress,floatfix]{revtex4-1}
\usepackage{caption}
\usepackage{graphicx} % needed for figures
\usepackage{subfigure}  
\usepackage{dcolumn}   % needed for some tables
\usepackage{bm}  % for math
\usepackage{epstopdf} 
\usepackage{amsmath}      
\usepackage{amssymb}   % for math
\usepackage{bbold}
\usepackage{slashed}
\usepackage{array}
\usepackage{booktabs}
\begin{document}

\title{Yang-Mills Structure for Electron-Phonon Interactions}
\author{Jamie ~M.~Booth}
\email{jamie.booth@rmit.edu.au}

\author{Salvy ~P. ~Russo}

\affiliation{ARC Centre of Excellence in Exciton Science, RMIT University, Melbourne VIC 3001, Australia}

\affiliation{Theoretical and Chemical Quantum Physics, RMIT University, Melbourne VIC 3001, Australia}

\date{\today}

\begin{abstract}
This work presents a method of grouping the electron spinors and the acoustic phonon modes of polar crystals such as metal oxides into an SU(2) gauge theory. The gauge charge is the electron spin, which is assumed to couple to the transverse acoustic phonons on the basis of spin ordering phenomena in crystals such as V$_{2}$O$_{3}$ and VO$_{2}$, while the longitudinal mode is neutral. A generalization the Peierls mechanism is presented based on the discrete gauge invariance of crystals and the corresponding Ward-Takahashi identity. The introduction of a band index violates the Ward-Takahashi identity for interband transitions resulting in a longitudinal component appearing in the upper phonon band. Thus both the spinors and the vector bosons acquire mass and a crystal with an electronic band gap and optical phonon modes results. In the limit that the coupling of bosons charged under the SU(2) gauge group goes to zero, breaking the electron U(1) symmetry recovers the BCS mechanism. In the limit that the neutral boson decouples, a Cooper instability mediated by spin-wave exchange results from symmetry breaking, i.e. unconventional superconductivity mediated by magnetic interactions.
\end{abstract}

%\pacs{71.30.+h,71.27.+a,74.20.Pq,75.10.-b,71.20.-b}
\maketitle

\section{Introduction}
There currently exist a number of seemingly intractable problems in Condensed Matter physics (by intractable it is meant that some decades have passed since they were first identified without a solution being found). Mechanisms of metal-insulator transitions in metal oxides\cite{Imada1998,Takagi2010} and high temperature superconductivity in the cuprates and pnictides are two examples.\cite{Carlson2008,Keimer2015} In addition, the cooperative interplay of magnetism and lattice distortions has also been emphasized in layered transition-metal dichalcogenides.\cite{Qiao2017,Calandra2018,Pasquier2018}

Since the early years of the development of quantum mechanics, Condensed Matter physics has developed theories built around specific Hamiltonians which are limiting cases of the general behavior of condensed, non-relativistic quantum systems, such as the Heisenberg Ferro-/Antiferromagnet, the Hubbard Model\cite{Hubbard1963} and its simplification the $t-J$ model,\cite{Spaek2007} the perturbative Peierls Mechanism of metal-insulator transitions,\cite{Altland2006} and the BCS theory of superconductivity.\cite{Bardeen1957} 

The purpose behind this work is to explore whether such seemingly disparate phenomena can arise from a single, simple, underlying theory. In other words, can a theory of interacting particles be written down which unifies these models into some deeper structure in the same way that the Standard Model of Particle Physics unifies the Strong, Weak and Electromagnetic Forces. There are two approaches to exploring the physics of metal-oxide systems (or more generally polar crystals) that can be employed. One is to attempt to simplify the problem by using scalar fields (i.e. spinless Fermions; the Schroedinger Equation, and scalar displacement fields for the lattice excitations etc.), and then attempt to use more sophisticated techniques, such as holography, to explore the characteristics of the remaining degrees of freedom. Such an approach has proven very useful in some metal oxides,\cite{Donos2012} and is currently a very active area of research.

Another approach, and the one taken here, is to leave the spin degrees of freedom in the Fermion wavefunctions, and recognize that the lattice vibration modes in more than one dimension are described by polarization vectors that transform as, well, vectors, and attempt to describe experimental systems using a gauge theory of interacting spinors and vector bosons. This approach may seem at face value over-complicated, but there are some huge advantages to it. 

The first, and most obvious, is that by leaving the spin degrees of freedom in, and by the choice of an appropriate gauge field, magnetic excitations can be combined with charge fluctuations in a natural way (this is illustrated in detail below). Another advantage is that Yang-Mills theories contain boson-boson interactions naturally, and therefore can account for phonon-phonon scattering, which leads to another significant advantage: there are huge simplifications to the calculation of scattering amplitudes in Yang-Mills theories which have been developed since the work of Parke and Taylor in the 1980s.\cite{Parke1986} Therefore, by expressing the physics of crystal lattices in this way, all of the sophisticated techniques of modern amplitude methods \cite{Dixon2013,Arkani-Hamed2017} (e.g. BCFW recursion etc. \cite{Britto2005}) can be employed,  at least at low energies where an approximate Poincar{\'e} invariance holds (more on this later), work will need to be done to extend this to high momenta analytically.
%Put closing statement as footnote?

The work described here is concerned with symmetry-breaking, and in particular the formation of massive excitations from massless constituents, and is focussed on electron-lattice interactions. It is found that an SU(2) gauge theory in which the transverse phonons are charged under the gauge group and couple to the both the spins and electric charges of the electrons, while the longitudinal mode is neutral and induces electric charge fluctuations only contains the possibilities of: conventional and un-conventional superconductivity, paramagnet to ferromagnet or antiferromagnet transitions, and also a non-perturbative metal-insulator transition which includes spin ordering in which mass is generated from a mechanism similar to neutrino oscillations. 

\section{Spinors and the Weyl Equation}
In problems such as metal-insulator transitions and superconductivity we are interested in the behavior of the electrons (and to some extent the lattice), and in particular the electrons on- or close to the Fermi surface, which act as metallic excitations before symmetry-breaking. For example in the cuprates these are the $d_{x^{2}-y^{2}}$ states, and in vanadium dioxide the vanadium $d^{1}$ states. It is therefore natural to concentrate solely on these degrees of freedom, and consider tight-binding wavefunctions for the electrons comprised of atomic-like orbitals:
\begin{equation}
\psi_{n\mathbf{k}}(\mathbf{r}) = \sum_{j, \mathbf{k}} \phi_{j}(\mathbf{r-\mathbf{R}})e^{i\mathbf{k}\mathbf{R}}
\end{equation}
where $n$ is the band index, $\phi_{j}(\mathbf{r})$ labels the atomic-like orbitals which are summed over to give the position state wavefunctions in each unit cell, $\mathbf{R}$ labels the set of lattice vectors which describe the translational symmetry of the lattice, and $\mathbf{k}$ is the wavevector which describes the spatial variation in the wavefunction amplitude. 

Assuming a 2-dimensional Fermi surface, the bands which form such a surface can be linearized at $\mathbf{k} = \mathbf{k}_{F}$ the Fermi wavevector, i.e. $E_{\mathbf{k-k_{F}}} = c(\mathbf{k-k_{F}})$ and shifting $\mathbf{k} \rightarrow \mathbf{k-k_{F}}$ the states above describe electrons and holes on the Fermi surface. However, unlike the Standard Model, the coordinate system of a crystal has a specific orientation, so no Poincar{\'e} group exists. Of course, there will be a discrete rotational and translational symmetry of the crystal given by its space group, but in general this will not be of much use to us, as it is not a Lie Group and therefore the considerable machinery of Poincar{\'e} invariance cannot be applied.

Therefore, it is important to realize that momentum states in crystals are, in general, not related by a simple transformation. We can state that apart from rotations and translations of the space group, momenta are related by \textit{scattering} processes, not symmetry transformations. Of course, momenta which are related by scale transformations, i.e. boosts along directions in momentum space given by the direction of the momentum vectors: $\mathbf{k}\rightarrow \alpha\mathbf{k}$, are related. This may seem extremely restrictive, but it is actually a considerable simplification. We can treat each radial \textit{direction} in momentum space separately, and sum over them to give the total result. For effects restricted to the Fermi surface this is simply equivalent summing over each point on the Fermi surface.

There is one symmetry operation which will be of considerable use in this work, and that is that for the crystal systems under investigation in this work (and indeed for almost all crystal systems), an inversion centre exists. Therefore there \textit{is} a symmetry operation relating momenta $\mathbf{k}$ and $-\mathbf{k}$. It is straightforward to prove that these states satisfy the Weyl equation in 4-dimensional space-time, as for each individual pair we can rotate the coordinate axes such that $p = (\frac{p}{v_{F}},0,0,p)$ giving:
\begin{equation}
i\gamma^{\mu}\partial_{\mu}\psi_{\mathbf{a}} = \begin{pmatrix}0&0&p_{0}-p_{3}&0\\0&0&0&p_{0}+p_{3}\\p_{0}+p_{3}&0&0&0\\0&p_{0}-p_{3}&0&0\end{pmatrix}\begin{pmatrix}\psi^{1}_{L}\\\psi^{2}_{L}\\\psi^{1}_{R}\\\psi^{2}_{R}\end{pmatrix}
\label{Weyl}
\end{equation}
using 
\begin{equation}
\psi_{\mathbf{a}}=\begin{pmatrix}
\psi_{L}\\\psi_{R}
\end{pmatrix}
\end{equation}
where the left-handed ($L$) and right-handed ($R$) states correspond to the two opposite helicity solutions occurring for the up- and down spin degrees of freedom of the electrons, and $v_{F}$ has been set to unity. However, while this can be done for each \textit{pair} of 3-momenta, $\mathbf{k}$ and $-\mathbf{k}$, the lattice structure does not have Poincar{\'e} invariance. Therefore, to be able to compare different momentum states, we need a way of satisfying $E=v_{F}\lvert p\rvert$ and also equation (\ref{Weyl}). The simplest method of achieving this is to allow complex momenta, and indeed this is also the manner in which the violation of Poincar{\'e} invariance is handled in modern amplitude methods such as BCFW recursion \cite{Britto2005}. Of course, for a complete theory we need solutions for the left- and right-handed states, but first we need to determine how the lattice can influence how they vary from point-to-point across the lattice.

\section{Bosons, Lattice Fluctuations and Relativity}
In simple condensed matter systems such as the archetypal linear chain, the potential energy associated with lattice fluctuations is usually expressed as a function of the interatomic spacing. This gives the usual dispersion relation:
\begin{equation}
\omega(k) = \sqrt{\frac{K}{M}}\bigg\lvert \textrm{sin}\bigg(\frac{ka}{2}\bigg)\bigg\rvert
\end{equation}
for a monatomic linear chain, where $K$ is the force constant and $M$ is the atomic mass. However, for octahedrally coordinated metal ions in metal oxides, this simple potential energy expression is not valid. The restoring force is less dependent on the metal-metal interactions (except in some cases which will be explored below), and more dependent on the metal-oxygen interactions. However, computing the acoustic mode eigenvectors for a system such as vanadium dioxide reveals that the oxygen atoms are effectively static (see Supporting Information). Thus the modes consist of the metal atoms rattling around inside the cage of oxygen atoms. 

\begin{figure}[h!]
\subfigure{\includegraphics[width=0.6\columnwidth]{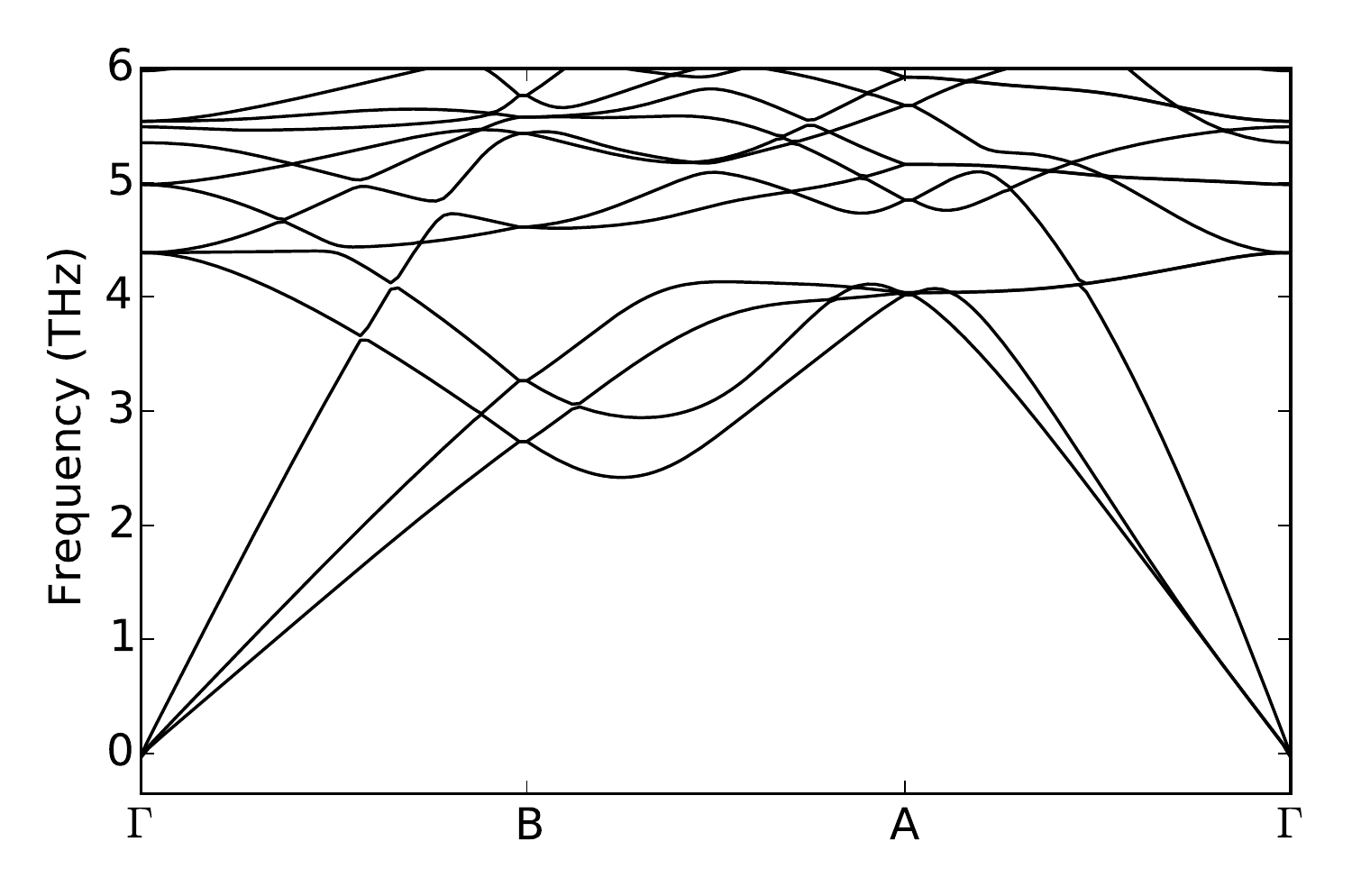}}{a)}
\subfigure{\includegraphics[width=0.6\columnwidth]{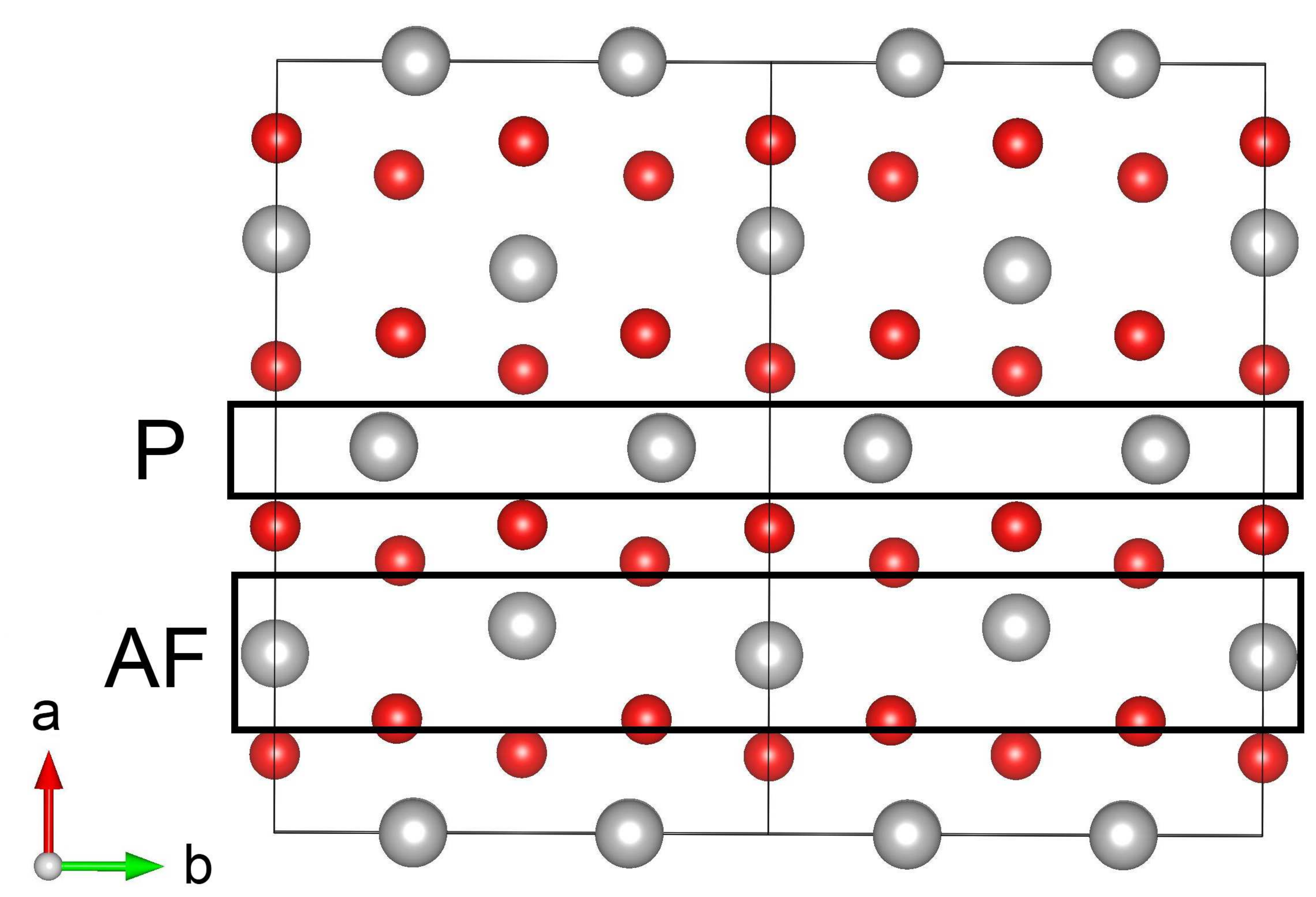}}{b)}\\
\subfigure{\includegraphics[width=0.4\columnwidth]{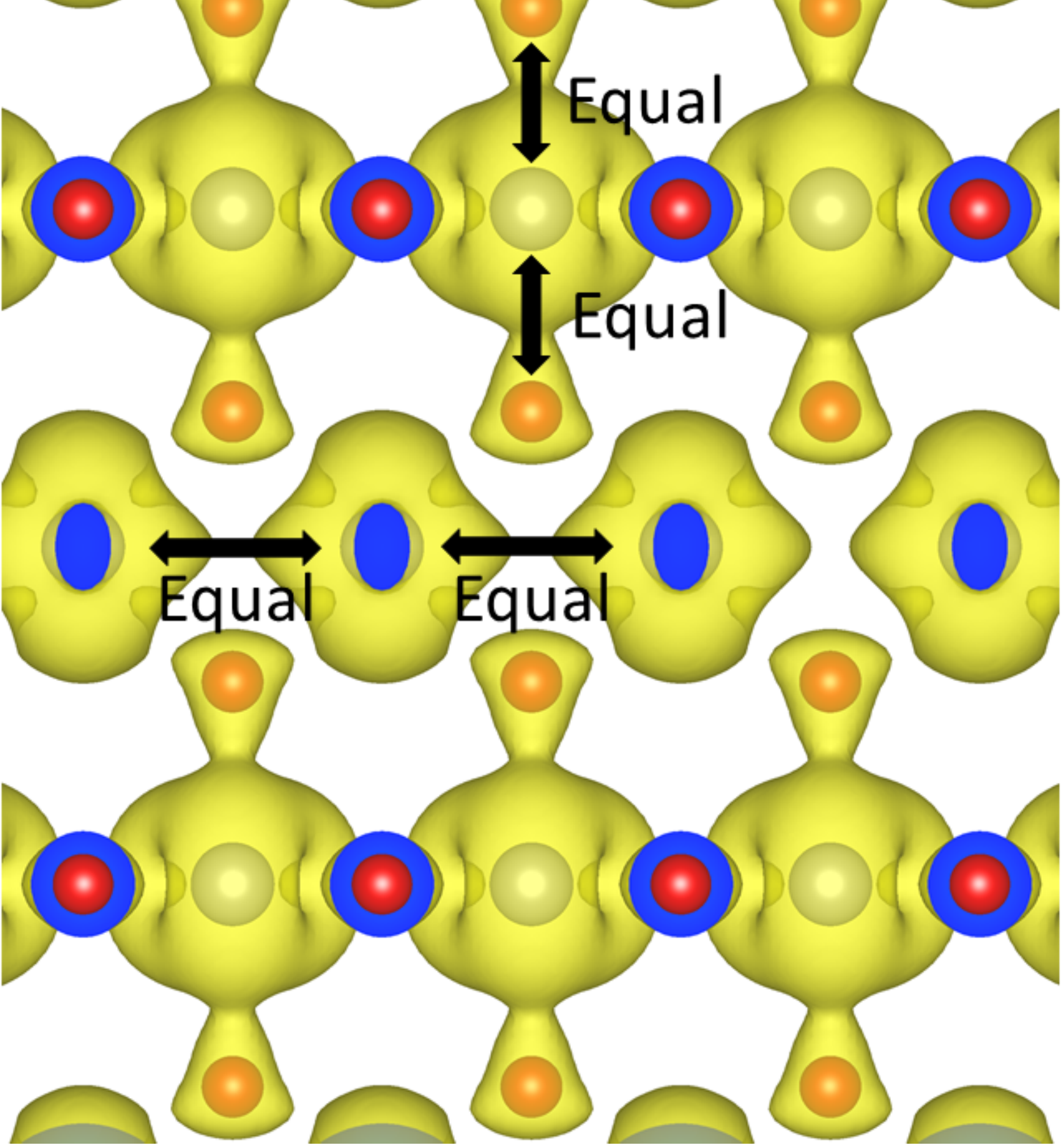}}{c)}
\subfigure{\includegraphics[width=0.4\columnwidth]{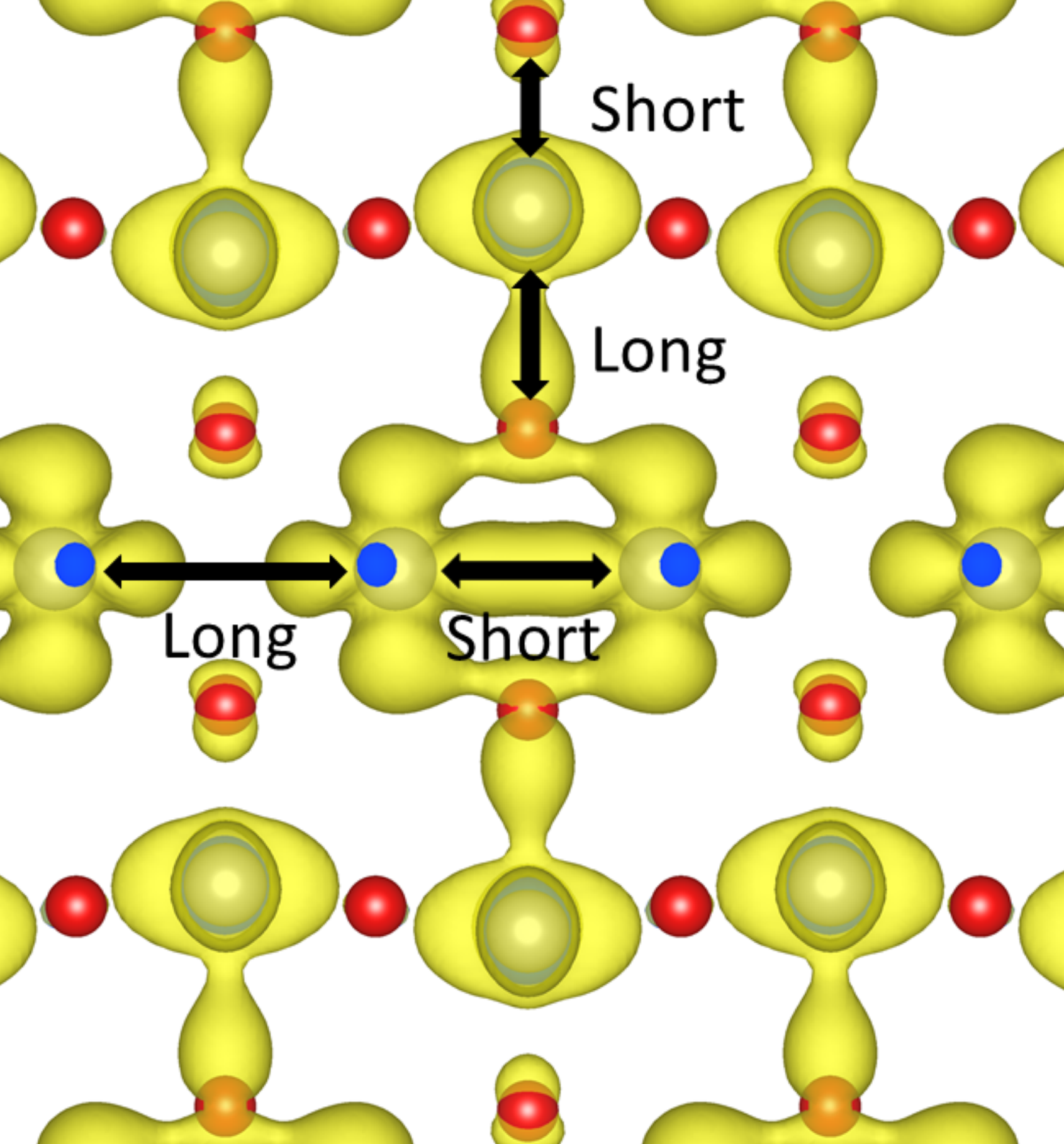}}{d)}
\caption{\raggedright{a) Crystal structure of the M$_{2}$ form of vanadium dioxide viewed down the monoclinic c-axis, where ``P" denotes the paired chain, while ``AF" denotes the antiferroelectrically distorted chain, b) Isosurface of the charge density of the tetragonal VO${2}$ structure, and c) Isosurface of the charge density of the M$_{2}$ VO$_{2}$ structure.}}
\label{M2}
\end{figure}

The questions then is, how do these modes disperse? For metallic systems, i.e. if the phonon modes are coupled to a Fermi surface, it is difficult to separate the restoring forces due to the repulsion of the electron orbitals, and the effect of electron-electron interactions. However, for the insulating, monoclinic form of vanadium dioxide, the $3d^{1}$ electrons on each metal atom are trapped in Peierls pairs and therefore coherent motion of the pairs (i.e. motion in which both paired atoms move in the same direction, corresponding to \textit{acoustic} modes) does not free up electrons to become itinerant (unlike an optical mode) and thus affect the total energy. Computing the phonon band structure (see Figure \ref{M2}a) of this form reveals something very interesting: the acoustic mode dispersion is approximately \textit{linear} almost all the way to the zone edge.

While small deviations are apparent, in this work we make the assumption that these are due to the phonon self-energy, and are small, which the dispersion relation indicates. From this linear dispersion it follows that for the acoustic modes, to leading order $p^{\mu}p_{\mu}=0$, where $p = (\frac{E_{\mathbf{p}}}{c},\mathbf{p})$; the 4-momentum, and $c$ is the proportionality constant to convert momentum to energy (the gradient). We then assume that the linearity is intrinsic to acoustic phonon modes in octahedrally coordinated metal oxides which are not coupled to a Fermi surface, and therefore assume that even in metallic structures the modes disperse linearly if the self-energy is neglected. This is a significant conjecture, which has equally significant consequences, and work is currently under way attempting to prove its validity. However, for the purposes of this work, we will take this conjecture to be valid, an explore its consequences. 

There is also another interesting phenomenon which occurs in metal oxide systems. Many systems exhibit crystal structure transformations in which charge and spin order, either simultaneously or separately\cite{Imada1998}. The aforementioned M$_{1}$ form of VO$_{2}$ is one such. At 340 K it undergoes a transition from a paramagnetic metal to an insulator in which the itinerant electrons form spin singlets while at the same time the tetragonal structure of the metal changes to a monoclinic form\cite{Morin1959,Eyert2002}. While this transition has significant potential, another closely related transition exhibits bevaviour which may provide a clue as to where the rich physics of transition metal oxides originates.
The M$_{2}$ form of vanadium dioxide\cite{Marezio1971} is also monoclinic, and like the M$_{1}$ form it also undergoes a metal-insulator transition, albeit at a slightly elevated temperature\cite{Booth2009}. However, the monoclinic structure has a particularly interesting feature, in that it is comprised of the same structural distortions as the M$_{1}$ form, but split across different vanadium chains. That is, in the M$_{1}$ form, all vanadium atoms pair up along the tetragonal c-axis, and at the same time undergo an antiferroelectric distortion, in which neighbouring metal ions displace in opposite directions along the long axis of the octahedron\cite{Marezio1971}. In the M$_{2}$ form these distortions occur on alternating chains, as Figure \ref{M2} illustrates. The key piece of information is that the antiferroelectrically distorted chain also orders antiferromagnetically, while the paired chain does not.\cite{Pouget1976} Similar cooperation between charge and spin ordering resulting in Mott physics has also been recently reported in 1T-NbSe$_{2}$, and in particular the Jahn-Teller distortion has a significant influence.\cite{Calandra2018}

\begin{figure}[h!]
\subfigure{\includegraphics[width=0.8\columnwidth]{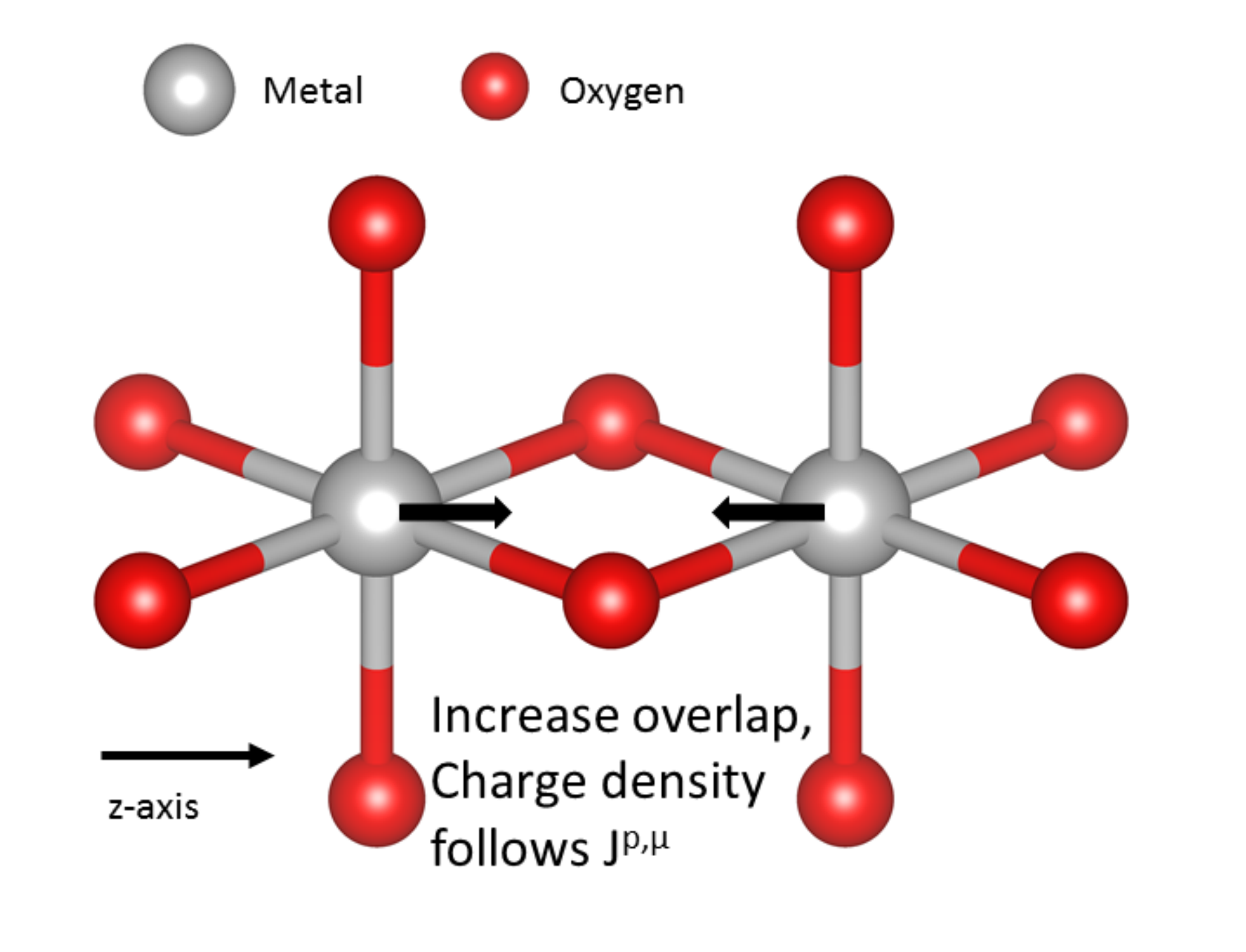}}{a)}
\subfigure{\includegraphics[width=0.9\columnwidth]{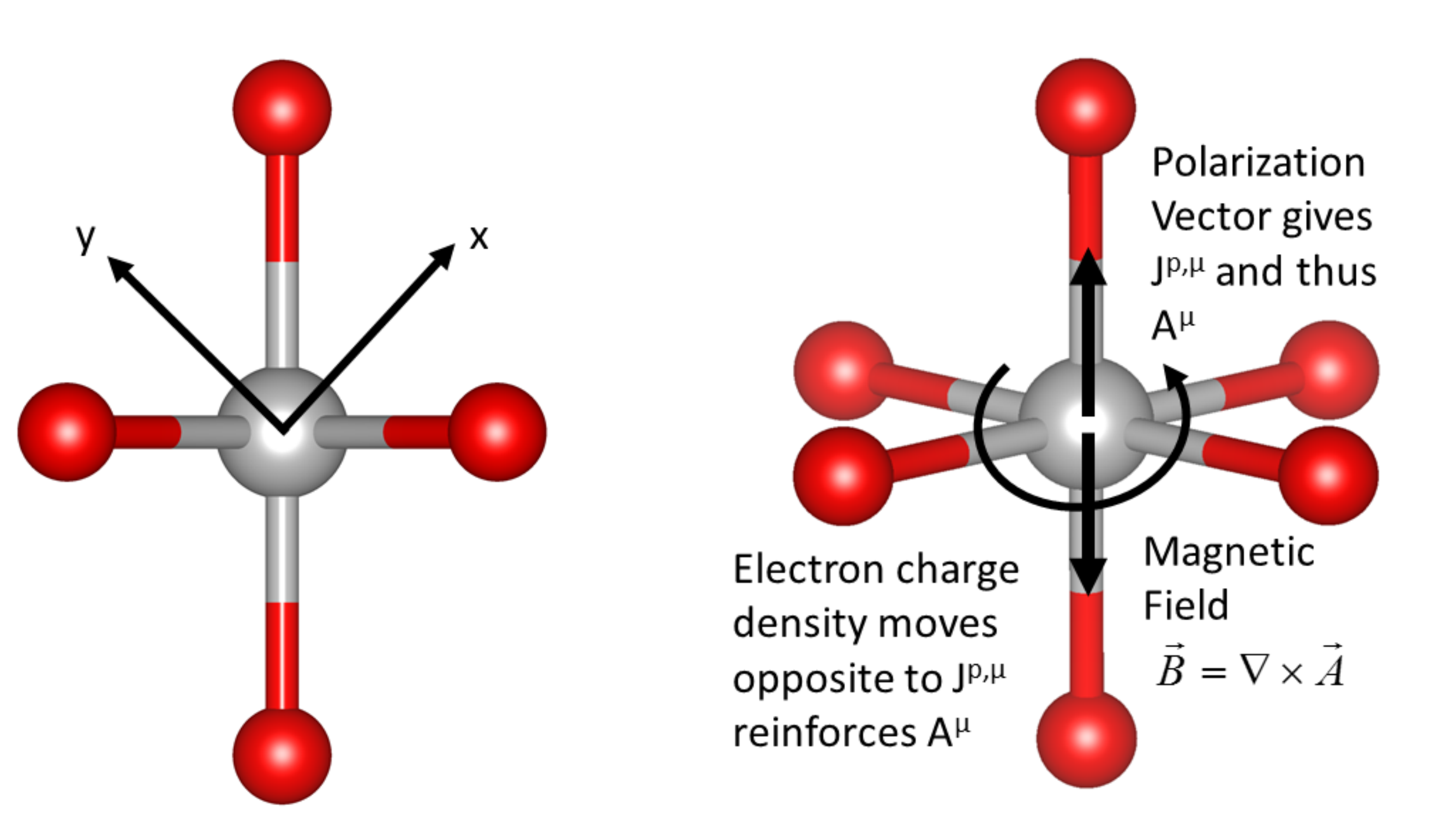}}{b)}
\caption{\raggedright{a) Schematic representation of the effect of the metal atom displacement, $J^{p,\mu}$ on the electron charge density for the pairing displacement, and b) representation of the reinforcement of $J^{p,\mu}$ by the electron charge density motion in the antiferroelectric distortion of M$_{2}$ VO$_{2}$.}}
\label{W}
\end{figure}

This suggests that the antiferromagnetic spin ordering is somehow related to the antiferroelectric distortion, while the pairing distortion has no effect on the spin. Looking a little closer at this, there is a possibly significant phenomenon occurring in the charge density when these two distortions manifest. Figures \ref{M2}b) and \ref{M2}c) present calculations of the charge density of the $d$-band electrons in tetragonal VO$_{2}$, and M$_{2}$ VO$_{2}$ respectively. The tetragonal structure exhibits charge density which is distributed equally between the vanadium and oxygen atoms, and does not accumulate in the inter-vanadium regions. However, the M$_{2}$ structure exhibits a very different charge ordering. From Fig \ref{M2}c) it is apparent that as the tetragonal structure transforms to the M$_{2}$ form, one of the chains dimerizes, and charge density accumulates between the paired vanadium atoms, indicated by the ``Short" label.

However, the antiferroelectrically distorted chain exhibits the opposite behaviour. The symmetric charge density of the tetragonal structure deforms such that more accumulates between the vanadium atom and one of the oxygens, however it accumulates in the \textit{long} inter-atomic spacing. This is most likely an effect of electrostatic repulsion, however it may have a significant effect. In the paired chain, the motion of the vanadium atom, which is partially positively charged, creates a positive current, which the electron charge density follows, and therefore the positive and negative currents will cancel each other. Figure \ref{W}a illustrates this schematically.

For the antiferroelectrically distorting chain however, the motion of the vanadium atom again sets up a positive current, however since the electron charge density moves in the opposite direction the current is reinforced, not cancelled, and a magnetic field results, as Fig \ref{W}b) illustrates. If the phonon mode responsible for this distortion is a zone edge mode of the tetragonal structure (there is significant evidence to suggest that this at least the case for M$_{1}$ VO$_{2}$ \cite{Budai2014}), neighbouring vanadium atoms will be moving in opposite directions and therefore the magnetic fields generated on each site will be in opposite directions, resulting in antiferromagnetic alignment of the 3$d^{1}$ electron spins. Here we make another significant conjecture: the different motions of the charge density for the pairing and antiferroelectric distortions correspond to coupling to charge only, and charge and spin respectively. As before, work is in progress to determine the validity of this assumption, but for now we assume that is \textit{is} valid, and explore the consequences. 

Therefore, it seems that there may be \textit{three} different effects of the metal atom motion on the localized electrons. There is a ``Neutral" phonon, which affects the charge density, but not the spin, and there are two ``Charged" bosons, which can align the spin up- or down. In VO$_{2}$ the neutral and charged bosons have orthogonal polarization vectors, and it is then natural to group the three boson fields into an SU(2) gauge theory, with the boson fields parametrized by the SU(2) generators. For the vector fields we can make the usual assumption that the solution to the field equation will be of the form \cite{Ashcroft1976}:
\begin{equation}
W^{a}_{\mu}(x)\sim \sum_{p}\epsilon_{\mu}(p)e^{ipx}
\label{W_approx}
\end{equation} 
where $\epsilon_{\mu}(p)$ is the polarization vector for each momentum state $p$, and using the almost-relativistic character of the boson dispersion, each boson can be quantized as per the usual procedure to give:
\begin{equation}
W_{\mu}^{a}(x) = \int \frac{d^{3}p}{{2\pi}^\frac{3}{2}2E_{\mathbf{p}}^{\frac{1}{2}}}\sum_{\lambda}\big[\hat{a}\epsilon^{\lambda}_{\mu}(p)e^{ip_{\mu}x^{\mu}} + \hat{a}^{\dagger}\epsilon^{*\lambda}_{\mu}(p)e^{-ip_{\mu}x^{\mu}}\big]
\label{W-field}
\end{equation}
where $\mu$ is a spacetime index running from $0 \rightarrow 3$, $a$ is a color index running from $1\rightarrow 3$, $x = (x^{0},x^{i}) = (t,\mathbf{x})$, and $\epsilon_{\mu}^{\lambda}(p)$ is the polarization vector as per equation (\ref{W_approx}), and we approximate the Brillouin Zone sum by an integral. These can be used to define an interaction vertex in which the charge density of the electrons couples to the vibrational modes in the usual manner. 

The question is now, what do these bosons act on? Given that experimentally metal-oxide systems exhibit spin and charge-ordering transitions, and superconductivity how do we group the spinors such that we can reproduce the experimentally observed behavior? The most obvious place to start is with BCS Theory. We might expect that BCS theory would drop out of the interaction vertex of a system in which the charged gauge coupling goes to zero, as this would be the limit in which the crystal is non-polar, corresponding to systems such as Al, or Nb.

This gives as possibilities:
\begin{equation}
\psi_{\mathbf{a}} = \begin{pmatrix}\hat{c}_{\mathbf{k}\uparrow}\\ \hat{c}^{\dagger}_{\mathbf{-k}\downarrow}\end{pmatrix}, \begin{pmatrix}\hat{c}_{\mathbf{-k}\downarrow}\\ \hat{c}^{\dagger}_{\mathbf{k}\uparrow}\end{pmatrix}, \begin{pmatrix}\hat{c}^{\dagger}_{\mathbf{k}\downarrow}\\ \hat{c}_{\mathbf{-k}\uparrow}\end{pmatrix}, \begin{pmatrix}\hat{c}^{\dagger}_{\mathbf{-k}\uparrow}\\ \hat{c}_{\mathbf{k}\downarrow}\end{pmatrix}
\end{equation}
where we have used the 3-vectors to label the momenta to express the helicities more clearly. Parametrizing the SU(2) interaction vertex in the usual manner:
\begin{equation}
\hat{W}_{\mu}=\sigma_{a}\cdot W^{a}_{\mu}
\end{equation}
where $a = 1, 2, 3$ is a color index, $\mu$ is a spacetime index and the $\sigma_{a}$ are the usual Pauli matrices:
\begin{equation}
\sigma_{1} = \begin{pmatrix}0&1\\1&0\end{pmatrix}; \sigma_{2} = \begin{pmatrix}0&-i\\i&0\end{pmatrix}; \sigma_{3} = \begin{pmatrix}1&0\\0&-1\end{pmatrix}
\end{equation}
therefore we have:
\begin{equation}
\hat{W}^{a}_{\mu}=\begin{bmatrix}W^{3}_{\mu}&W^{1}_{\mu}-iW^{2}_{\mu}\\W^{1}_{\mu}+iW^{2}_{\mu}&-W^{3}_{\mu}\end{bmatrix}
\end{equation}
We then expect electron phonon interactions to be of the obvious form:
\begin{multline}
g_{a}\bar{\psi}\gamma^{\mu}\hat{W}^{a}_{\mu}\psi=\\g_{(1,2,3)}\begin{pmatrix}\bar{\psi}_{\mathbf{a^{\prime}}},\bar{\psi}_{\mathbf{b}^{\prime}}\end{pmatrix}\gamma^{\mu}\begin{pmatrix}W^{3}_{\mu}&W^{1}_{\mu}-iW^{2}_{\mu}\\W^{1}_{\mu}+iW^{2}_{\mu}&-W^{3}_{\mu}\end{pmatrix}\begin{pmatrix}\psi_{\mathbf{a}} \\ \psi_{\mathbf{b}}\end{pmatrix}
\end{multline}
where the gamma matrices are expressed (in the chiral basis) in two-component form as:
\begin{equation}
\gamma^{0} = \begin{pmatrix}0&\mathbb{1}\\\mathbb{1}&0\end{pmatrix}, \gamma^{i} = \begin{pmatrix}0&\sigma^{i}\\-\sigma^{i}&0\end{pmatrix}
\end{equation}
and thus
\begin{equation}
\bar{\psi} = \psi^{\dagger}\gamma^{0} = (\bar{\psi}_{\mathbf{a}},\bar{\psi}_{\mathbf{b}}) = (\psi^{\dagger}_{\mathbf{a}}\gamma^{0},\psi^{\dagger}_{\mathbf{b}}\gamma^{0})
\end{equation}

\section{Diagonal Interactions}
Setting $F^{1}_{\mu\nu}=F^{2}_{\mu\nu} = 0$ for the field strength tensors gives:
\begin{multline}
\begin{pmatrix}\bar{\psi}_{\mathbf{a^{\prime}}},\bar{\psi}_{\mathbf{b}^{\prime}}\end{pmatrix}g_{3}\gamma^{\mu}\begin{pmatrix}W^{3}_{\mu}&0\\0&-W^{3}_{\mu}\end{pmatrix}\begin{pmatrix}\psi_{\mathbf{a}}\\\psi_{\mathbf{b}}\end{pmatrix}\\=\bar{\psi}_{\mathbf{a^{\prime}}}g_{3}\gamma^{\mu}W^{3}_{\mu}\psi_{\mathbf{a}}-\bar{\psi}_{\mathbf{b^{\prime}}}g_{3}\gamma^{\mu}W^{3}_{\mu}\psi_{\mathbf{b}}
\end{multline}
where $i = 1, 2, 3$. These are the familiar matrix elements of a standard Abelian gauge theory, which represent the traditional electron-phonon interaction involved in for example the BCS theory of superconductivity, with the exception that the Yang-Mills field strength tensor $F^{3}_{\mu\nu}$ contains a quadratic term which gives self-interactions. In the language of differential forms: $F^{3}_{\mu\nu} = dF^{3}_{\mu} + F^{1}_{\mu}\wedge F^{2}_{\nu}$. For conventional electron-phonon interactions, for example in monovalent metals which are non-polar crystals, $F^{1}_{\mu\nu}=F^{2}_{\mu\nu} = 0$ and the quadratic term vanishes, giving the standard Abelian Field Strength Tensor. In this respect, the $W^{3}_{\mu}$ boson is like the neutral boson of the weak interaction, it carries zero angular momentum.

\section{Off-Diagonal Terms}
So far, so familiar. However, in order to contain gauge ``charge" coupling the off-diagonal terms contain spin raising and lowering operators. To see how the these arise, we set $W^{3}_{\mu}=0$ for clarity and expand the interaction for $\mu = 1,2$ to get:
\begin{multline}
\begin{pmatrix}\bar{\psi}_{\mathbf{a^{\prime}}},\bar{\psi}_{\mathbf{b}^{\prime}}\end{pmatrix}g_{(1,2)}\gamma^{\mu}\begin{pmatrix}0&W^{1}_{\mu}-iW^{2}_{\mu}\\W^{1}_{\mu}+iW^{2}_{\mu}&0\end{pmatrix}\begin{pmatrix}\psi_{\mathbf{a}}\\\psi_{\mathbf{b}}\end{pmatrix}=\\
\begin{pmatrix}\bar{\psi}_{\mathbf{a^{\prime}}},\bar{\psi}_{\mathbf{b}^{\prime}}\end{pmatrix}g_{(1,2)}\begin{pmatrix}0&\gamma^{1}\\\gamma^{1}&0\end{pmatrix}\begin{pmatrix}0&W^{1}_{1}-iW^{2}_{1}\\W^{1}_{1}+iW^{2}_{1}&0\end{pmatrix}\begin{pmatrix}\psi_{\mathbf{a}}\\\psi_{\mathbf{b}}\end{pmatrix}\\
+\begin{pmatrix}\bar{\psi}_{\mathbf{a^{\prime}}},\bar{\psi}_{\mathbf{b}^{\prime}}\end{pmatrix}g_{(1,2)}\begin{pmatrix}0&\gamma^{2}\\\gamma^{2}&0\end{pmatrix}\begin{pmatrix}0&W^{1}_{2}-iW^{2}_{2}\\W^{1}_{2}+iW^{2}_{2}&0\end{pmatrix}\begin{pmatrix}\psi_{\mathbf{a}}\\\psi_{\mathbf{b}}\end{pmatrix}\\
\end{multline}
Setting $g_{1}W^{1}_{1} = g_{2}W^{2}_{2}$ to illustrate this most clearly we get a term:
\begin{equation}
\begin{pmatrix}\bar{\psi}_{\mathbf{a^{\prime}}},\bar{\psi}_{\mathbf{b}^{\prime}}\end{pmatrix}\begin{pmatrix}0&g_{1}W^{1}_{1}(\gamma^{1}-i\gamma^{2})\\g_{1}W^{1}_{1}(\gamma^{1}+i\gamma^{2})&0\end{pmatrix}\begin{pmatrix}\psi_{\mathbf{a}}\\\psi_{\mathbf{b}}\end{pmatrix}
\end{equation}
If both $\psi_{\mathbf{a}}$ and $\psi_{\mathbf{b}}$ are in eigenstates of $S_{Z}$, and remembering:
\begin{equation}
\gamma^{i} = \begin{pmatrix}0&\sigma^{i}\\-\sigma^{i}&0\end{pmatrix}
\end{equation}
this gives the familiar spin raising and lowering operators, $S^{+}=\sigma^{1}+i\sigma^{2}$, and $S^{-}=\sigma^{1}-i\sigma^{2}$:
\begin{equation}
\bar{\psi}_{\mathbf{a^{\prime}}}g_{1}W^{1}_{1}\begin{pmatrix}0&\hat{S}^{-}\\-\hat{S}^{-}&0\end{pmatrix}\psi_{\mathbf{b}}+\bar{\psi}_{\mathbf{b^{\prime}}}g_{1}W^{1}_{1}\begin{pmatrix}0&\hat{S}^{+}\\-\hat{S}^{+}&0\end{pmatrix}\\\psi_{\mathbf{a}}
\label{Spin_Operators}
\end{equation}
with the negative sign in the $\gamma^{i}$ accounting for the opposite helicities of the two-component spinors in each four-component spinor such that the Weyl equation for each is satisfied.

\begin{figure}[h!]
\subfigure{\includegraphics[width=0.5\columnwidth]{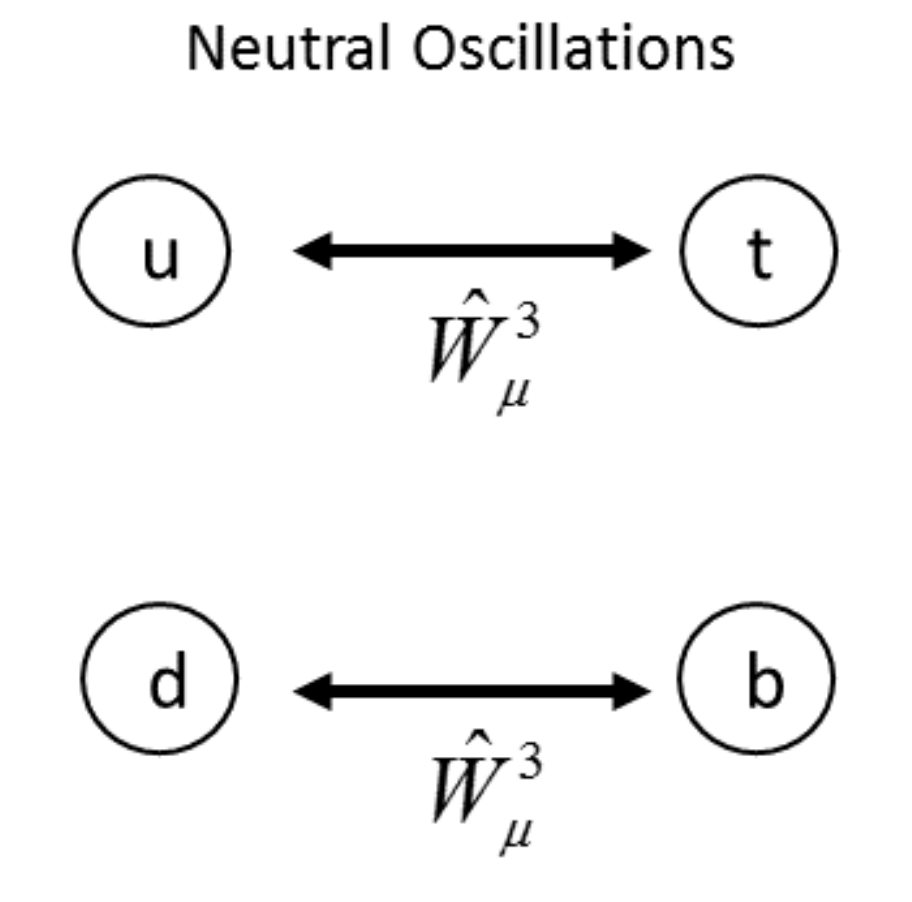}}{a)}\\
\subfigure{\includegraphics[width=0.55\columnwidth]{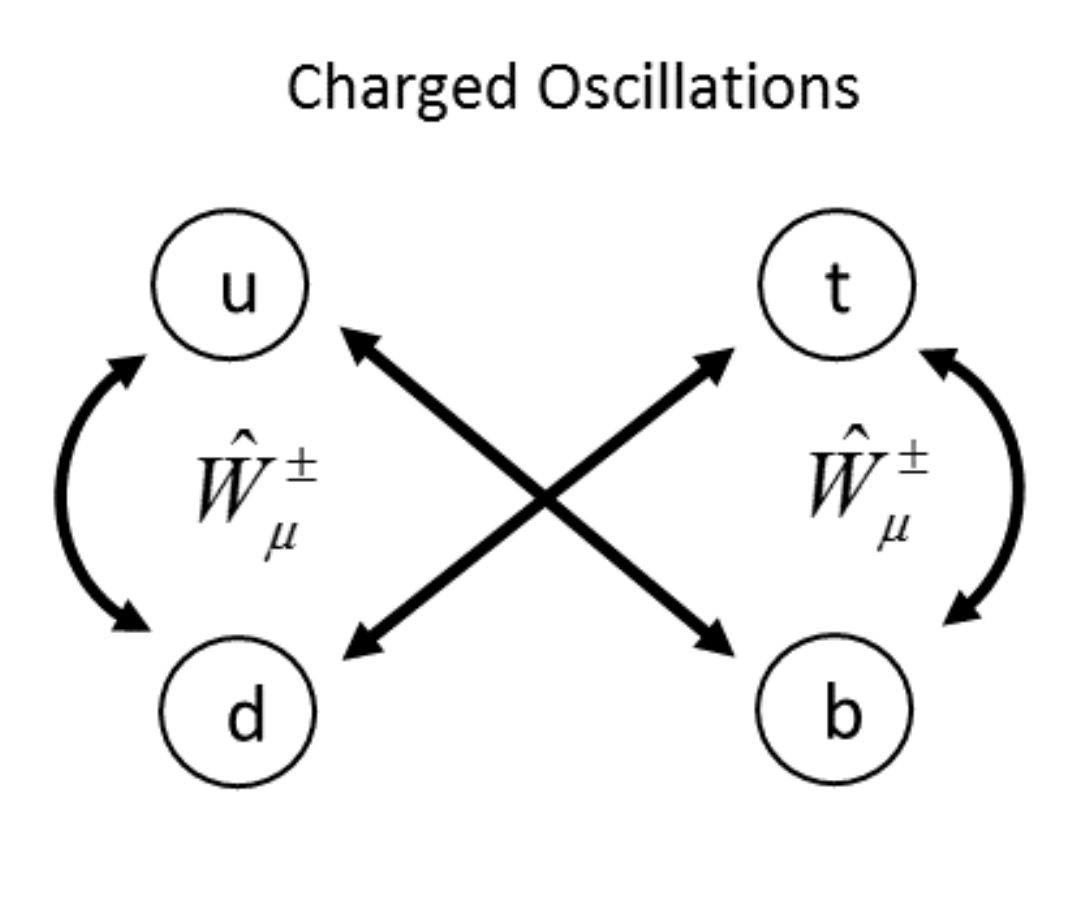}}{b)}
\caption{\raggedright{Schematic representation of the transformations enacted by the a) Neutral boson $W^{3}_{\mu}$ and b) the Charged bosons $W^{\pm}_{\mu}$}.}
\label{Schematics}
\end{figure}

This provides us with an easy way to determine how to group the Nambu spinors. If the interaction vertex is rewritten:
\begin{equation}
\begin{pmatrix}\bar{\psi}_{\mathbf{a^{\prime}}},\bar{\psi}_{\mathbf{b}^{\prime}}\end{pmatrix}g_{(+,-,3)}\gamma^{\mu}\begin{pmatrix}W^{3}_{\mu}&W^{+}_{\mu}\\W^{-}_{\mu}&-W^{3}_{\mu}\end{pmatrix}\begin{pmatrix}\psi_{\mathbf{a}} \\ \psi_{\mathbf{b}}\end{pmatrix}
\end{equation}
where $W^{\pm}_{\mu} = W^{1}_{\mu}\pm W^{2}_{\mu}$ then we group the spin down electrons and spin up holes into $\psi_{\mathbf{a}}$, and the spin up electrons and spin down holes into $\psi_{\mathbf{b}}$. Thus the $W^{\pm}_{\mu}$ describe transformations between spinors which contain electrons of opposite momentum and spin if $W^{1}_{\mu}$ and $W^{2}_{\mu}$ are in phase (this can easily be generalized into arbitrary charge density relationships, which will be explored in the context of vanadium dioxide later). However, there are four Nambu spinors, and therefore there are two each of the $\psi_{\mathbf{a}}$ and $\psi_{\mathbf{b}}$. We can therefore group the spinors into flavours, and generations.
\begin{multline}
\begin{pmatrix}\hat{c}^{\dagger}_{\mathbf{k}\uparrow}\\\hat{c}_{\mathbf{-k}\downarrow}\end{pmatrix} = \textrm{up},
\begin{pmatrix}\hat{c}^{\dagger}_{\mathbf{-k}\downarrow}\\\hat{c}_{\mathbf{k}\uparrow}\end{pmatrix} = \textrm{down},\\
\begin{pmatrix}\hat{c}_{\mathbf{k}\downarrow}\\\hat{c}^{\dagger}_{\mathbf{-k}\uparrow}\end{pmatrix} = \textrm{top},
\begin{pmatrix}\hat{c}_{\mathbf{-k}\uparrow}\\\hat{c}^{\dagger}_{\mathbf{k}\downarrow}\end{pmatrix} = \textrm{bottom}
\end{multline}
The naming convention follows the spin of the holes in each Dirac spinor, which is done to preserve the commutation relations of the Pauli matrices. This can be summarized in table form as:
\begin{center}
\begin{tabular}{c|cc}
\toprule
Flavour &  \multicolumn{2}{c}{Generation}\\
  & 1   & 2 \\
  \hline
\textbf{a} & up   & top \\
\textbf{b} & down   & bottom \\
\bottomrule
\end{tabular}
\end{center}
Thus action of the phonons on the grouped Dirac spinors in all its gory detail becomes:
\begin{multline}
g_{(+,-,3)}\gamma^{\mu}\hat{W}_{\mu}(x)\psi=\begin{pmatrix}\gamma^{\mu}W^{3}_{\mu}(x)& \gamma^{\mu}W^{-}_{\mu}(x)\\ & \\ \gamma^{\mu}W^{+}_{\mu}(x)&-\gamma^{\mu}W^{3}_{\mu}(x)\end{pmatrix}\begin{pmatrix} \hat{c}^{\dagger}_{\mathbf{k}\uparrow}\\\hat{c}_{\mathbf{-k}\downarrow}\\\hat{c}^{\dagger}_{\mathbf{-k}\downarrow}\\\hat{c}_{\mathbf{k}\uparrow}\end{pmatrix}\\ g_{(+,-,3)}\gamma^{\mu}\hat{W}_{\mu}(x)\psi=\begin{pmatrix}\gamma^{\mu}W^{3}_{\mu}(x)&\gamma^{\mu}W^{-}_{\mu}(x)\\ & \\  \gamma^{\mu}W^{+}_{\mu}(x)&-\gamma^{\mu}W^{3}_{\mu}(x)\end{pmatrix}\begin{pmatrix} \hat{c}_{\mathbf{k}\downarrow}\\\hat{c}^{\dagger}_{\mathbf{-k}\uparrow}\\\hat{c}_{\mathbf{-k}\uparrow}\\\hat{c}^{\dagger}_{\mathbf{k}\downarrow}\end{pmatrix}
\end{multline}
Therefore $W^{+}_{\mu}(x)$ can scatter: a bottom to a top, a down to an up, and a bottom to an up (with zero wavevector) etc., and so on. A schematic of the transformations the bosons perform is presented in Fig (\ref{Schematics})
\section{Mass Generation}
\subsection{Spinor Mass from Neutral Oscillations}
Defining the chirality operator in the usual way, we can redefine the spinors as per:
\begin{equation}
\frac{1}{2}(1-\gamma^{5})\psi = \begin{pmatrix}\psi_{L}\\0\end{pmatrix} = \psi_{L} \quad\textrm{and}\quad \frac{1}{2}(1+\gamma^{5})\psi = \begin{pmatrix}0\\\psi_{R}\end{pmatrix} = \psi_{R}
\end{equation}
Thus:
\begin{equation}
\psi^{1,2}_{\mathbf{a},\mathbf{b}} = \psi_{L}+\psi_{R}
\end{equation}
and identifying the upper and lower components of each 4-component spinor (i.e. the $\psi_{\mathbf{a}}$ and $\psi_{\mathbf{b}}$) as left- and right-handed chiral spinors (it is straightforward to prove these satisfy the Weyl equation in metallic systems), i.e.
\begin{equation}
\begin{pmatrix}\hat{c}^{\dagger}_{\mathbf{k}\uparrow}\\\hat{c}_{\mathbf{-k}\downarrow}\end{pmatrix} = \begin{pmatrix}u_{L}\\u_{R}\end{pmatrix}\quad \textrm{and}\quad \begin{pmatrix}\hat{c}_{\mathbf{k}\downarrow}\\\hat{c}^{\dagger}_{\mathbf{-k}\uparrow}\end{pmatrix} = \begin{pmatrix}t_{L}\\t_{R}\end{pmatrix}
\end{equation}
and
\begin{equation}
\begin{pmatrix}\hat{c}^{\dagger}_{\mathbf{-k}\downarrow}\\\hat{c}_{\mathbf{k}\uparrow}\end{pmatrix} = \begin{pmatrix}d_{L}\\d_{R}\end{pmatrix}\quad \textrm{and}\quad \begin{pmatrix}\hat{c}_{\mathbf{-k}\uparrow}\\\hat{c}^{\dagger}_{\mathbf{k}\downarrow}\end{pmatrix} = \begin{pmatrix}b_{L}\\b_{R}\end{pmatrix}
\end{equation}
we can see how the phonon field gaining a Vacuum Expectation Value (VEV) can result in massive spinors in the same manner as neutrino oscillations (i.e. the Rabi cycle). While all of the $W^{a}_{\mu}$ contribute to lattice potential fluctuations, let us focus on $W^{3}_{\mu}$ for clarity, and note that a longitudinal phonon with wavevector $\pm2\mathbf{k}$ will scatter $u_{L}\rightarrow t_{R}$, $u_{R}\rightarrow t_{L}$, $d_{L}\rightarrow b_{R}$ and $d_{R}\rightarrow b_{L}$ and vice versa, where $\mathbf{k}$ is the wavevector of the electron state. So, giving $W^{3}_{\mu}$ a VEV ($\epsilon^{\lambda}_{0}(p) = 0$) with wavevector $2\mathbf{k}$ we get:
\begin{equation}
W^{3}_{\mu}(x)\rightarrow \langle W^{3}_{0}\rangle+W^{3}_{\mu}(x)
\end{equation}
To maintain the spin ordering, i.e. to give neutral oscillations as per Figure \ref{Schematics}, there will be constraints on the polarization vector. Looking at the interaction of the boson with an incoming spinor such as $t_{L}$ (dropping the coupling constant and the outgoing spinor to see the interaction more clearly):
\begin{multline}
ig_{3}\bar{\psi}\gamma^{\mu}\hat{W}^{3}_{\mu}\psi\rightarrow\bar{\sigma}^{\mu}\epsilon_{\mu}(p)e^{ipx}\begin{pmatrix}0\\1\end{pmatrix}e^{ikx}=\\\begin{pmatrix}\epsilon_{0}-\epsilon_{3}&-(\epsilon_{1}-i\epsilon_{2})\\-(\epsilon_{1}+i\epsilon_{2})&\epsilon_{0}+\epsilon_{3}\end{pmatrix}\begin{pmatrix}0\\1\end{pmatrix}e^{i(p+k)x}\\=\begin{pmatrix}-\epsilon_{1}+i\epsilon_{2}\\\epsilon_{0}+\epsilon_{3}\end{pmatrix}e^{i(p+k)x}
\end{multline}
where $\bar{\sigma}=(\mathbb{1},-\sigma^{i})$. By giving the $\hat{W}^{3}_{\mu}(x)$ field a VEV, and setting $\epsilon_{0} = 0$, to maintain the spin orientation we can have:
\begin{equation}
\epsilon_{1}=i\epsilon_{2},\quad\textrm{or}\quad\epsilon_{1}=\epsilon_{2}=0,\quad\textrm{with}\quad\epsilon_{3}=1
\end{equation}
Choosing the easy path and defining the orientation of the polarization vector as being down the z-axis ($\epsilon_{\mu}(p)=(0,0,0,1)$) the full interaction vertex; $-g_{3}\bar{\psi}\gamma^{\mu}\hat{W}_{\mu}\psi$ gives:
\begin{equation}
g_{3}\langle W^{3}_{3}\rangle\bar{u}_{R}t_{L} + g_{3}\langle W^{3}_{3}\rangle\bar{u}_{L}t_{R}\quad\textrm{\dots}\quad
\label{DiracMass} 
\end{equation}
where $\bar{\psi}=\psi^{\dagger}\gamma^{0}$ and we have switched to the Dirac representation, i.e.
\begin{equation}
\gamma^{0}=\begin{pmatrix}
1&0&0&0\\
0&1&0&0\\
0&0&-1&0\\
0&0&0&-1
\end{pmatrix}
\end{equation}
This is identical to the Peierls metal-insulator transition, where the system becomes unstable to to a potential with wavelength $2\mathbf{k}_{F}$, however this is now a \textit{three dimensional} mechanism, the pairing wavevector can point in any direction, but the \textit{polarization} vector points down the z-axis. 

\begin{figure}[h!]
\subfigure{\includegraphics[width=0.5\columnwidth]{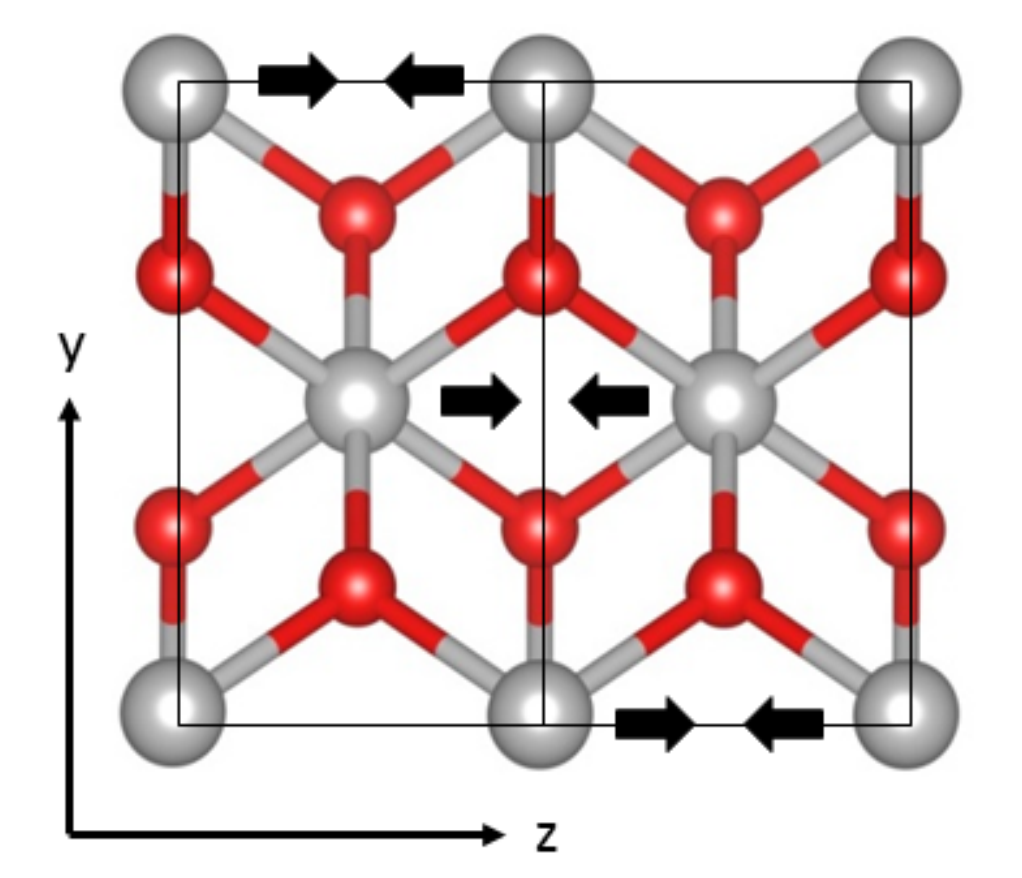}}{a)}\\
\subfigure{\includegraphics[width=0.5\columnwidth]{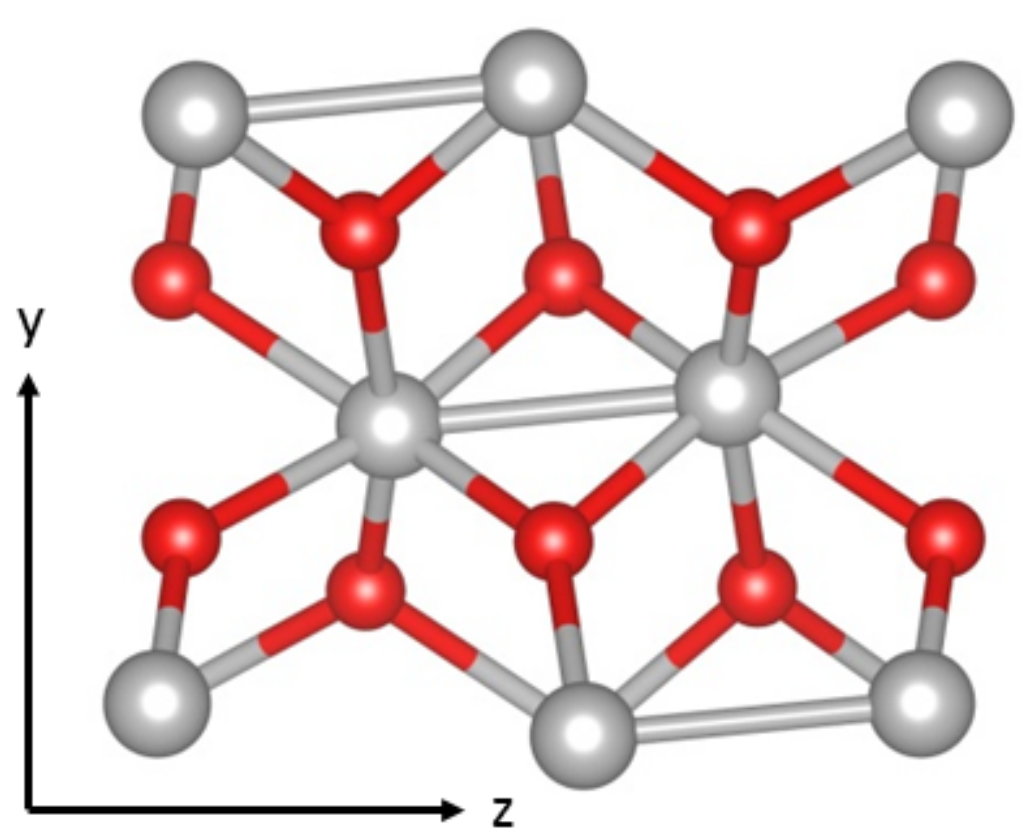}}{b)}
\caption{\raggedright{a) Tetragonal structure of VO$_{2}$ with the z-axis oriented down the crystallographic c-axis showing the pairing distortion of the MIT, unit cell boundaries are marked in black and b) view of the M$_{1}$ VO$_{2}$ structure showing the Peierls pairs and their relative positions on different vanadium chains.}}
\label{VO2_Pairing}
\end{figure}

To give this some context, the aforementioned metal-insulator transition of M$_{1}$ VO$_{2}$ contains just such a Peierls pairing component (there is also an interplay of spin degrees of freedom and strong correlations which we will not deal with here). Figure \ref{VO2_Pairing} illustrates the polarization vectors of the pairing distortion (we ignore the antiferroelectric distortion for now), and as vanadium atoms from neighbouring unit cells are moving towards each other, this defines a zone edge mode, however the pairing displacements also have a non-zero component in the y-direction; the vanadium atoms are moving in opposite directions in neighbouring unit cells along the y-axis, and thus the wavevector of this phonon mode has two non-zero components, $\mathbf{k}_{y}$ and $\mathbf{k}_{z}$. However, the polarization vector only has one component.

Equivalently, we can take $u_{L}$ and $t_{R}$, and grouping them together gives the Hamiltonian:
\begin{equation}
\psi^{\dagger}\hat{H}\psi = \begin{pmatrix}u^{\dagger}_{R}\quad t^{\dagger}_{L}\end{pmatrix}\begin{pmatrix}\epsilon_{-\mathbf{k}}&\langle W^{3}_{3}\rangle\\\langle W^{3}_{3}\rangle&\epsilon_{\mathbf{k}}\end{pmatrix}\begin{pmatrix}u_{R}&t_{L}\end{pmatrix}
\label{PeierlsH}
\end{equation}
Diagonalizing gives as eigenvectors the linear combinations:
\begin{equation}
|\psi_{+}\rangle = |u_{R}\rangle + |t_{L}\rangle,\quad \textrm{and}\quad |\psi_{-}\rangle = |u_{R}\rangle - |t_{L}\rangle
\end{equation}
with eigenvalues $E_{+} = \epsilon_{\mathbf{k}}+\langle W^{3}_{3}\rangle$ and $E_{-} = \epsilon_{\mathbf{k}}-\langle W^{3}_{3}\rangle$ assuming that $\epsilon_{\mathbf{k}}=\epsilon_{-\mathbf{k}}$, i.e. both states sit on the Fermi surface. Time evolving, and computing the probability of transitioning from $t_{L}$ to $u_{R}$ in the usual Rabi fashion gives:
\begin{equation}
P_{L\rightarrow R}(t) = \textrm{sin}^{2}\bigg({\frac{(E_{+}-E_{-})}{2\hbar}}t\bigg)
\end{equation}
Thus the probability of an electron being in either a left- or right-handed state is oscillatory in time, with a frequency given by the magnitude of the phonon VEV: $(E_{+}-E_{-}) = \langle W^{3}_{3}\rangle$. This is precisely the same statement as the ``mass" terms in equation (\ref{DiracMass}) above, generated from the phonon VEV taking left-handed particles into right-handed and vice versa.

There is also the option of breaking the symmetry with the $\hat{W}^{\pm}_{\mu}$, however some consideration reveals that giving these phonons a VEV does not result in a ground state with fluctuating spins. From Figure (\ref{Schematics}), and using the $u_{R}$ spinor as an example we see that the $\hat{W}^{+}_{\mu}$ can scatter: $u_{R}\rightarrow d_{R}$ (which is not a Dirac mass), or $u_{R}\rightarrow b_{L}$. However, for both processes, giving $\hat{W}^{+}_{\mu}$ a VEV will decouple it from the electron spin. Reiterating:
\begin{equation}
W^{a}_{\mu}(x)\sim \sum_{p}\epsilon_{\mu}(p)e^{ipx}
\end{equation}
this oscillating polarization vector creates a positive current density $J^{p,\mu} = (\rho,\mathbf{J})$, however the time derivative, or the energy $\hbar\omega$ goes to zero as the phonon gains a VEV (remembering that $e^{ipx} = \textrm{exp}(\frac{i\omega_{\mathbf{p}}}{c}-i\mathbf{p}.\mathbf{x})$ as the phonons disperse linearly). Therefore the time-dependence of the phonon vanishes, and therefore the current, and thus the associated magnetic field. Thus, spin ordering is a dynamic process which will occur before the phonon VEV sets in, i.e. above T$_{c}$, and below T$_{c}$ oscillations of the type described by equation (\ref{DiracMass}) which either flip helicity by flipping the spin, or preserve helicity by flipping the spin and the momentum, will not be present in the ground state. Of course spin fluctuations due to the $\hat{W}^{\pm}_{\mu\nu}$ propagators can still manifest, and a charged boson VEV will still couple to the electron charge (i.e. the U(1) gauge charge), but not the SU(2) gauge charge.

\begin{figure}[h!]
\subfigure{\includegraphics[width=0.7\columnwidth]{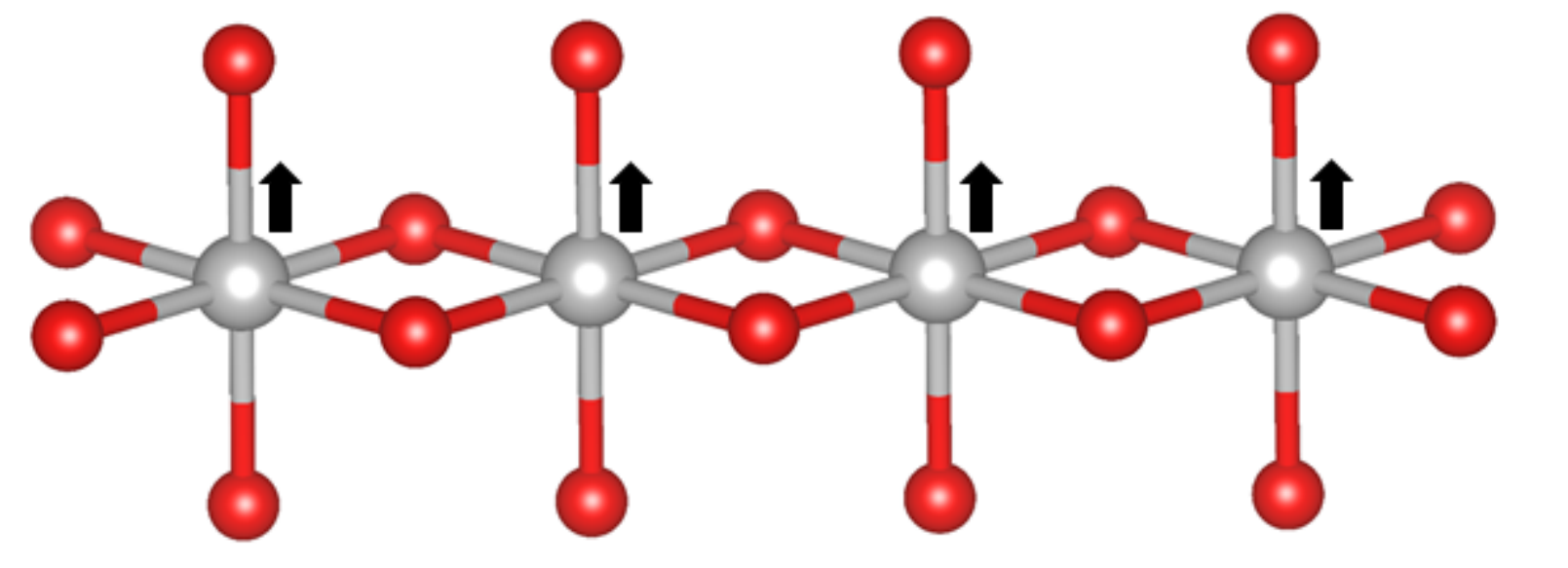}}{a)}\\
\subfigure{\includegraphics[width=0.7\columnwidth]{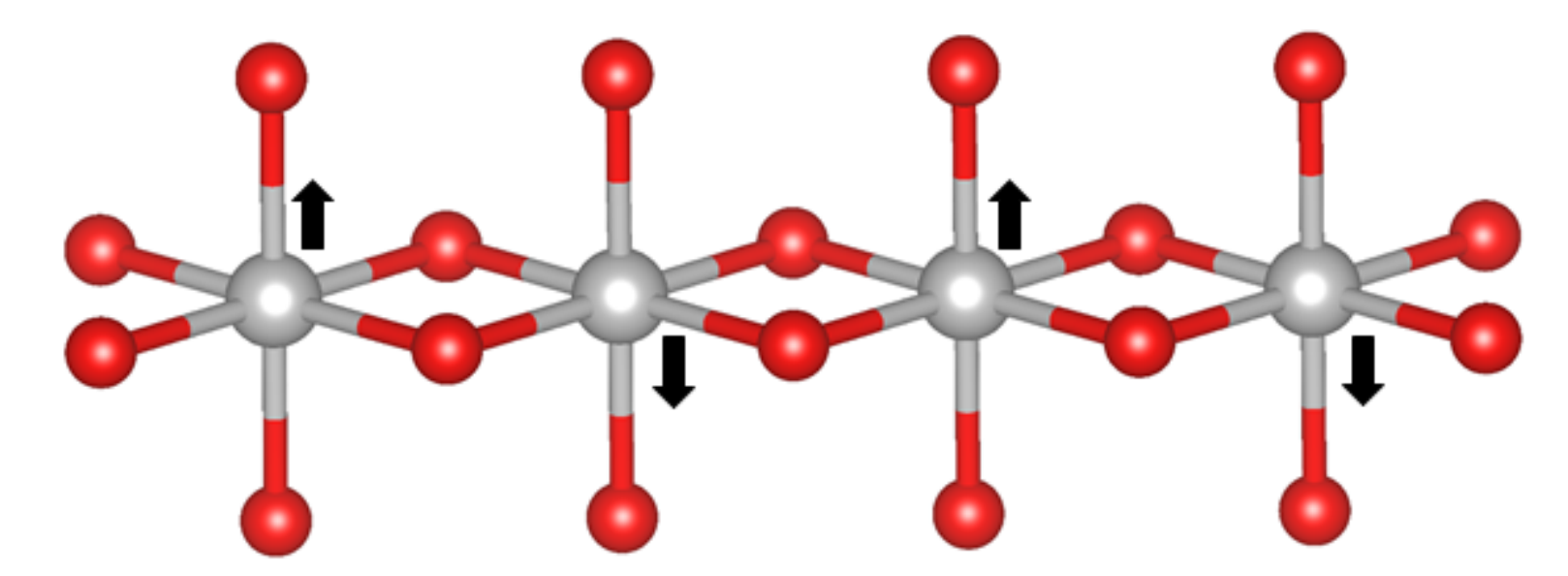}}{b)}
\caption{\raggedright{a) Ferroelectric distortion of a chain of octahedrally coordinated metal atoms, b) Antiferroelectric distortion of a chain of octahedrally coordinated metal atoms.}}
\label{Spin_Ordering}
\end{figure}

\subsection{Spin Ordering}
While it is straightforward to define the spin raising and lowering operators as per equation \ref{Spin_Operators}, there is a slight additional subtlety to their implementation. If the direction of the magnetic field is dependent on the direction of the current, then describing the phonons by normal modes which oscillate as a function of time means that the current will also oscillate, and the magnetic field will change direction. Therefore, if the spin coupling is via the magnetic field, the spin operators themselves will oscillate between raising and lowering as a function of time, for example at spacetime point $x$ (i.e. a particular unit cell at a particular time) we might have:
\begin{multline}
\gamma^{\mu}\hat{W}^{a}_{\mu}(x)\psi=\\\gamma^{\mu}\begin{pmatrix}0&\hat{W}^{1}_{\mu}(x)-i\hat{W}^{2}_{\mu}(x)\\\hat{W}^{1}_{\mu}(x)+i\hat{W}^{2}_{\mu}(x)&0\end{pmatrix}\begin{pmatrix}\psi_{\mathbf{a}}\\\psi_{\mathbf{b}}\end{pmatrix}
\end{multline}
Simplifying this by writing the polarization vectors for $\hat{W}^{1}_{\mu}$ and $\hat{W}^{2}_{\mu}$ as $\epsilon^{1}_{\mu}=(0,1,0,0)$ and $\epsilon^{2}_{\mu}=(0,0,i,0)$ respectively, this just gives spin raising operators acting on $\psi_{\mathbf{b}}$ and lowering operators acting on $\psi_{\mathbf{a}}$ (remembering that the $\gamma^{\mu}$ contain two Pauli matrices in the off-diagonal positions). For the filled states, since the $\psi_{\mathbf{a}}$ and $\psi_{\mathbf{b}}$ are defined to be in eigenstates of $\hat{S}_{z}$, this will flip the spins corresponding to a transition between the generations of spinors.

However, time evolving to spacetime point ($x^{\prime}$), where ($x^{\prime}-x)=(\frac{1}{2}T,0,0,0)$, i.e just half the period of the oscillation gives:
\begin{multline}
\gamma^{\mu}\hat{W}^{a}_{\mu}(x^{\prime})\psi=\\\gamma^{\mu}\begin{pmatrix}0&-\hat{W}^{1}_{\mu}(x^{\prime})+i\hat{W}^{2}_{\mu}(x^{\prime})\\-\hat{W}^{1}_{\mu}(x^{\prime})-i\hat{W}^{2}_{\mu}(x^{\prime})&0\end{pmatrix}\begin{pmatrix}\psi_{\mathbf{a}}\\\psi_{\mathbf{b}}\end{pmatrix}\\=\begin{pmatrix}0&-\hat{S}^{-}(x^{\prime})\\-\hat{S}^{+}(x^{\prime})&0\end{pmatrix}\begin{pmatrix}\psi_{\mathbf{a}}\\\psi_{\mathbf{b}}\end{pmatrix}
\end{multline}
This means that when the current oscillates in the opposite direction the amplitude to flip the spin is not zero, this is unphysical, as is the fact that both operators are acting at the same spacetime point, however if we phase shift $\hat{W}^{2}_{\mu}$ with respect to $\hat{W}^{1}_{\mu}$ for either the raising or lowering operator we recover the correct physics. Taking $\omega(W^{1})=2\omega(W^{2})$, we must add a phase shift to the $\hat{W}^{2}$ part of $\hat{W}^{+}_{\mu}$ to cancel the spin operator corresponding to the wrong direction of the current:
\begin{multline}
\gamma^{\mu}\hat{W}^{a}_{\mu}(x)\psi=\\\gamma^{\mu}\begin{pmatrix}0&\hat{W}^{1}_{\mu}(x)-i\hat{W}^{2}_{\mu}(x)\\\hat{W}^{1}_{\mu}(x)+i\hat{W}^{2}_{\mu}(x)e^{i\pi}&0\end{pmatrix}\begin{pmatrix}\psi_{\mathbf{a}}\\\psi_{\mathbf{b}}\end{pmatrix}\\
=\gamma^{\mu}\begin{pmatrix}0&\hat{W}^{1}_{\mu}(x)-i\hat{W}^{2}_{\mu}(x)\\\hat{W}^{1}_{\mu}(x)-i\hat{W}^{2}_{\mu}(x)&0\end{pmatrix}\begin{pmatrix}\psi_{\mathbf{a}}\\\psi_{\mathbf{b}}\end{pmatrix}
\end{multline}
The $\gamma^{\mu}(\hat{W}^{1}_{\mu}-i\hat{W}^{2}_{\mu})$ term on the bottom row acts on (filled) eigenstates of $\hat{S}_{z}$ which correspond to down spins, and will therefore return zero. Time evolving for $\frac{1}{2}T(\hat{W}^{2})$ will give:
\begin{multline}
\gamma^{\mu}\hat{W}^{a}_{\mu}(x^{\prime})\psi=\\\gamma^{\mu}\begin{pmatrix}0&\hat{W}^{1}_{\mu}(x^{\prime})+i\hat{W}^{2}_{\mu}(x^{\prime})\\\hat{W}^{1}_{\mu}(x^{\prime})+i\hat{W}^{2}_{\mu}(x^{\prime})&0\end{pmatrix}\begin{pmatrix}\psi_{\mathbf{a}}\\\psi_{\mathbf{b}}\end{pmatrix}
\end{multline}
and the $\gamma^{\mu}(\hat{W}^{1}_{\mu}+i\hat{W}^{2}_{\mu})$ term on the top will act on an up eigenstate of $\hat{S}_{z}$ and return zero, while the raising operator on the bottom row will give a spin flip from down to up.

To see how the interaction vertex can order spins along a chain of metal atoms and referring to Figure \ref{Spin_Ordering}a, if the black arrows correspond to the instantaneous direction of the polarization vector (which creates a magnetic field of a specific orientation) at each metal atom, then this field will cause the spins to order in the same direction, and time-evolving will just give the fields oscillating, but all oscillating in phase, and thus the spin ordering will be ferromagnetic. This would be a zone centre mode, as each atom is oscillating in the same direction, with the same amplitude, and thus the wavelength is infinite ($\mathbf{k}=0$). 

Figure \ref{Spin_Ordering}b describes a zone edge mode, in which neigbouring metal atoms experience magnetic fields oscillating in opposite directions due to the out of phase oscillations of the polarization vectors, thus the wavelength of such a mode is 2$\mathbf{a}$, or twice the lattice spacing (i.e. $\mathbf{k=\frac{\pi}{\mathbf{a}}}$), if each octahedral cluster corresponds to one unit cell. This will order the spins antiferromagnetically, they will still oscillate from up-to-down, but 180 $^\circ$ out of phase. This is the type of ``frozen phonon" seen to correspond to antiferromagnetic ordering in compounds such as M$_{1}$ and M$_{2}$ VO$_{2}$ as a result of their structural phase transitions. If such ordering were to occur from coherent oscillations just above T$_{c}$, along with the symmetry-breaking of equation(\ref{DiracMass}) at T$_{c}$ in the tetragonal structure of VO$_{2}$, then we might expect the formation of localized singlets on the paired vanadium atoms, along with the transition to monoclinic symmetry from the pairing and antiferroelectric distortion VEVs, echoing what is seen in experiment.

\subsection{Superconductivity and Varying the Coupling}
The approach can be taken to pair states. Given that the electron and hole spinors have been grouped into forms which preserve the Nambu form, superconductivity arises naturally, with the Nambu Hamiltonian:
\begin{equation}
\hat{H} = \begin{pmatrix}
\xi_{\mathbf{k}}&-\Delta\\-\bar{\Delta}&-\xi_{-\mathbf{k}}
\end{pmatrix}
\end{equation}

where $\Delta = g\langle \Omega|\hat{c}_{\mathbf{k}\uparrow}\hat{c}_{-\mathbf{k}\downarrow}|\Omega\rangle$, being diagonalized by the usual Bogoliubov procedure. Conventional superconductivity is mediated by the exchange of neutral bosons, $W^{3}_{\mu}(x)$, which produce electron density waves, and this gives the traditional BCS version of superconductivity. In terms of mass generation, noting that the electron (and hole) states of each ``generation" form a Cooper pair:

\begin{equation}
\begin{pmatrix}\hat{c}_{\mathbf{-k}\downarrow}\\\hat{c}_{\mathbf{k}\uparrow}\end{pmatrix}=\begin{pmatrix}u_{R}\\d_{R}\end{pmatrix}=\psi_{R}\quad\textrm{and}\quad \psi_{L} = \begin{pmatrix}\hat{c}_{\mathbf{k}\downarrow}\\\hat{c}_{-\mathbf{k}\uparrow}\end{pmatrix} = \begin{pmatrix}t_{L}\\b_{L}\end{pmatrix}
\end{equation}
we can insert the pair VEV into the interaction vertex, and set the charged boson couplings to zero (i.e. $g_{1}=g_{2} = 0$):
\begin{equation}
-g^{2}_{3}\bar{\psi}\gamma^{\mu}\psi\hat{W}^{3}_{\mu\nu}\bar{\psi}\gamma^{\nu}\psi \sim -\Delta\bar{\psi}_{L}\psi_{R}-\Delta\bar{\psi}_{R}\psi_{L}
\label{SC}
\end{equation}
giving a mass term for the Lagrangian in the usual manner.

However, there are also the charged bosons, i.e. the propagator in equation (\ref{SC}) could be either $W^{+}_{\mu\nu}$ or $W^{-}_{\mu\nu}$ which can also induce attractive interactions between electrons in the same manner (i.e. if the propagators are off-shell). Thus there exists another, unconventional, superconductivity, which occurs via both charge- and spin fluctuations.

We can now explore the influence that the gauge couplings $g_{(+,-,3)}$ have on the behavior of the crystal. If we start with a system in which the couplings are approximately equal, then the bare interaction vertices correspond to approximately the same amplitudes, and we might expect both charge and spin fluctuations to occur simultaneously. If we take such a system and change it such that the couplings $g_{+,-}$ become smaller, we might then expect that only charge ordering phenomena will manifest. Going the other way, and doping the structure with holes or extra electrons will destabilize the pairing interactions with respect to the formation of static charge density waves, however while the occurrence of phonon-mediated spin waves will be only slightly affected, as the restoring force has an elastic (electrostatic) component, as well as the spin coupling and therefore will not be disrupted as significantly by the presence of holes, or extra electrons, and we might therefore expect this behavior to dominate in this regime. 

This competition between the $g_{3}$ and $g_{+-}$ couplings and their sensitivities to doping may result in regions of the oxide phase diagram in which the charge and spin fluctuations manifest separately, with possibilities for exotic behavior, reminiscent of the phase diagram of the cuprate superconductors.\cite{Carlson2008}

\subsection{Lattices, Gauge Boson Mass and the Ward-Takahashi Identity}
In high energy physics the Ward-Takahashi identity reflects the unphysical nature of the gauge redundancy, and reveals that the longitudinal components of massless vector bosons decouple from scattering amplitudes.\cite{Peskin2016} By defining a discrete version of the Ward-Takahashi identity, we can see how the symmetry-breaking of Equation (\ref{DiracMass}) reflects the emergence of a massive boson.

The propagator for a massive spin 1 boson in the Unitary gauge is given by:
\begin{equation}
W^{a}_{\mu\nu} = \frac{i}{k^{2}-m^{2}+i\epsilon}\bigg(g_{\mu\nu}-\frac{k_{\mu}k_{\nu}}{m^{2}}\bigg)
\end{equation}
The term: 
\begin{equation}
\frac{k_{\mu}k_{\nu}}{m^{2}}
\end{equation}
is the longitudinal component, which for massless bosons disappears. In continuous systems this is a trivial manifestation of the Ward-Takahashi identity, while for discrete systems we can see that this will occur for boson momenta which coincide with the reciprocal lattice vectors. 

Approximating the full scattering vertex $-igW^{a}_{\mu\nu}\bar{\psi}\gamma^{\nu}\psi$ with $\sim -ig\slashed{k}_{\nu}\bar{\psi}\gamma^{\nu}\psi$ to see the effect most clearly gives:
\begin{multline}
\frac{i}{(\slashed{p}+\slashed{k})-m+i\epsilon}(-ig\slashed{k})\frac{i}{\slashed{p}-m+i\epsilon} =\\ ig\bigg(\frac{1}{\slashed{p}-m+i\epsilon}-\frac{1}{\slashed{p}-m+\slashed{k}+i\epsilon}\bigg) = 0
\label{WT}
\end{multline}
where we have used $\slashed{k}=(\slashed{p}+\slashed{k})-\slashed{p}$ from momentum conservation at the vertex. This will vanish identically if $k = G$ since $p = p+G$ in a discrete system, and thus since the propagator is the Green function of the field equation of motion, and $p$ and $p+G$ give identical behaviour, the propagators cancel. However, if we consider the case of the symmetry-breaking represented by the Hamiltonian of equation (\ref{PeierlsH}), the diagonalization process gives two states which correspond to the same wavevector. Thus there is now a band index associated with the electronic states $|\psi_{+}\rangle$ and $|\psi_{-}\rangle$. In the parlance of condensed matter physics, we have $|\psi_{n\mathbf{p}}\rangle$, where the index $n$ denotes which eigenfunction we are considering. Thus after symmetry-breaking $|\psi_{n\mathbf{p}}\rangle\neq |\psi_{n^{\prime}\mathbf{p}}\rangle$, and therefore the identity of equation (\ref{WT}) is \textit{not} satisfied for inter-band scattering (i.e. $n\rightarrow n^{\prime}$). The longitudinal component of the propagator does not vanish: the boson has acquired a mass. This component \textit{does} still vanish for intra-band transitions (i.e. $n = n^{\prime}$), and therefore there are both massive and massless phonons: the optical and acoustic branches. 

Since:
\begin{equation}
\bar{\psi}\slashed{k}\psi = \bar{\psi}\gamma^{\mu}\partial_{\mu}\psi = \partial_{\mu}J^{\mu}
\end{equation}
we note that we note that current conservation, $\partial_{\mu}J^{\mu} = 0$, is violated for inter-band transitions and thus they give non-vanishing electric currents. There is an excitation energy associated with forming currents which corresponds to the phonon VEV ($\langle W^{3}_{\mu}\rangle$); the system is insulating.

\section{Conclusions}
By conjecturing that the almost-linear dispersion of acoustic phonon modes in M$_{1}$ VO$_{2}$ is a result of the fact that the restoring force in metal oxide crystals in the absence of interactions with the electron field depends most significantly on repulsion of the metal atom by the oxygen ligands, the acoustic modes due to their linear dispersions are able to be represented by relativistic vector bosons. Assuming also that the anomalous charge density motion of transverse phonons which correspond to oscillations along the Jahn-Teller axis of the octahedra produces a magnetic field which couples to electron spins, and combining this with the above assumption allows an SU(2) Yang-Mills theory of electron-phonon interactions to be postulated. 

This approach contains a wealth of interesting behavior due to the inclusion of the spinor and vector degrees of freedom. Such behavior includes: crystal structure transformations via spin- and charge-ordering, conventional superconductivity, and an unconventional superconductivity mediated by phonons which couple to the spin degrees of freedom of the electrons.

\section{Methods}
\subsection{Calculations}
The GW calculations were performed using the implementation of Shishkin and Kresse \cite{Shishkin2006,Shishkin2007} as contained in the Vienna Ab Initio Simulation Package (VASP),\cite{Kresse1996} after first calculating input wavefunctions using DFT\cite{Kohn1965} with GGA\cite{Perdew1996} functionals, on $8\times8\times6$ and  $4\times6\times6$ Monkhorst-Pack\cite{Monkhorst1976} \textbf{k}-space grids for the Tetragonal and M$_{2}$ structures respectively, using the Brillouin zone integration approach of Bloechl \textit{et al}.\cite{Bloechl1994} 

To determine the phonon band structure of M$_{1}$ VO$_{2}$, the monoclinic\cite{Longo1970} structural parameters were input to DFT geometry relaxations using the VASP code\cite{Kresse1996} and the Generalized Gradient Approximation to exchange and correlation of Perdew et al.,\cite{Perdew1996} on a 6$\times$6$\times$6 Monkhorst-Pack\cite{Monkhorst1976} k-space grid. The structure was then relaxed to the ground state using Methfessel and Paxton smearing\cite{Methfessel1989} and the conjugate gradient algorithm. The Hessian matrix of a 2$\times$2$\times$2 supercell was determined using Density Functional Perturbation Theory.\cite{Gonze1995b} The program Phonopy\cite{Togo2015} was used to calculate the phonon dispersion curves.

\section{Acknowledgements}
This work was supported by computational resources provided by the Australian Government through the National Computational Infrastructure under the National Computational Merit Allocation Scheme. The authors acknowledge the support of the ARC Centre of Excellence in Exciton Science (CE170100026), and useful discussions with S. Todd and S. Bilson-Thompson.
\section{Correspondence}
Correspondence and requests for materials should be addressed to JMB, email: jamie.booth@rmit.edu.au
\bibliography{C:/Local_Disk/GWApproximation/Bibliography/library}

\begin{thebibliography}{35}%
\makeatletter
\providecommand \@ifxundefined [1]{%
 \@ifx{#1\undefined}
}%
\providecommand \@ifnum [1]{%
 \ifnum #1\expandafter \@firstoftwo
 \else \expandafter \@secondoftwo
 \fi
}%
\providecommand \@ifx [1]{%
 \ifx #1\expandafter \@firstoftwo
 \else \expandafter \@secondoftwo
 \fi
}%
\providecommand \natexlab [1]{#1}%
\providecommand \enquote  [1]{``#1''}%
\providecommand \bibnamefont  [1]{#1}%
\providecommand \bibfnamefont [1]{#1}%
\providecommand \citenamefont [1]{#1}%
\providecommand \href@noop [0]{\@secondoftwo}%
\providecommand \href [0]{\begingroup \@sanitize@url \@href}%
\providecommand \@href[1]{\@@startlink{#1}\@@href}%
\providecommand \@@href[1]{\endgroup#1\@@endlink}%
\providecommand \@sanitize@url [0]{\catcode `\\12\catcode `\$12\catcode
  `\&12\catcode `\#12\catcode `\^12\catcode `\_12\catcode `\%12\relax}%
\providecommand \@@startlink[1]{}%
\providecommand \@@endlink[0]{}%
\providecommand \url  [0]{\begingroup\@sanitize@url \@url }%
\providecommand \@url [1]{\endgroup\@href {#1}{\urlprefix }}%
\providecommand \urlprefix  [0]{URL }%
\providecommand \Eprint [0]{\href }%
\providecommand \doibase [0]{http://dx.doi.org/}%
\providecommand \selectlanguage [0]{\@gobble}%
\providecommand \bibinfo  [0]{\@secondoftwo}%
\providecommand \bibfield  [0]{\@secondoftwo}%
\providecommand \translation [1]{[#1]}%
\providecommand \BibitemOpen [0]{}%
\providecommand \bibitemStop [0]{}%
\providecommand \bibitemNoStop [0]{.\EOS\space}%
\providecommand \EOS [0]{\spacefactor3000\relax}%
\providecommand \BibitemShut  [1]{\csname bibitem#1\endcsname}%
\let\auto@bib@innerbib\@empty
%</preamble>
\bibitem [{\citenamefont {Imada}\ \emph {et~al.}(1998)\citenamefont {Imada},
  \citenamefont {Fujimori},\ and\ \citenamefont {Tokura}}]{Imada1998}%
  \BibitemOpen
  \bibfield  {author} {\bibinfo {author} {\bibfnamefont {M.}~\bibnamefont
  {Imada}}, \bibinfo {author} {\bibfnamefont {A.}~\bibnamefont {Fujimori}}, \
  and\ \bibinfo {author} {\bibfnamefont {Y.}~\bibnamefont {Tokura}},\ }\href
  {\doibase 10.1103/RevModPhys.70.1039} {\bibfield  {journal} {\bibinfo
  {journal} {Rev. Mod. Phys.}\ }\textbf {\bibinfo {volume} {70}},\ \bibinfo
  {pages} {1039} (\bibinfo {year} {1998})}\BibitemShut {NoStop}%
\bibitem [{\citenamefont {Takagi}\ and\ \citenamefont
  {Hwang}(2010)}]{Takagi2010}%
  \BibitemOpen
  \bibfield  {author} {\bibinfo {author} {\bibfnamefont {H.}~\bibnamefont
  {Takagi}}\ and\ \bibinfo {author} {\bibfnamefont {H.~Y.}\ \bibnamefont
  {Hwang}},\ }\href {\doibase 10.1126/science.1182541} {\bibfield  {journal}
  {\bibinfo  {journal} {Science (80).}\ }\textbf {\bibinfo {volume} {327}},\
  \bibinfo {pages} {1601} (\bibinfo {year} {2010})}\BibitemShut {NoStop}%
\bibitem [{\citenamefont {Carlson}\ \emph {et~al.}(2008)\citenamefont
  {Carlson}, \citenamefont {Emery}, \citenamefont {Kivelson},\ and\
  \citenamefont {Orgad}}]{Carlson2008}%
  \BibitemOpen
  \bibfield  {author} {\bibinfo {author} {\bibfnamefont {E.~W.}\ \bibnamefont
  {Carlson}}, \bibinfo {author} {\bibfnamefont {V.~J.}\ \bibnamefont {Emery}},
  \bibinfo {author} {\bibfnamefont {S.~A.}\ \bibnamefont {Kivelson}}, \ and\
  \bibinfo {author} {\bibfnamefont {D.}~\bibnamefont {Orgad}},\ }\href
  {\doibase 10.1007/978-3-540-73253-2_21} {\bibfield  {journal} {\bibinfo
  {journal} {arXiv}\ ,\ \bibinfo {pages} {180}} (\bibinfo {year} {2008})},\
  \Eprint {http://arxiv.org/abs/0206217} {arXiv:0206217 [cond-mat]}
  \BibitemShut {NoStop}%
\bibitem [{\citenamefont {Keimer}\ \emph {et~al.}(2015)\citenamefont {Keimer},
  \citenamefont {Kivelson}, \citenamefont {Norman}, \citenamefont {Uchida},\
  and\ \citenamefont {Zaanen}}]{Keimer2015}%
  \BibitemOpen
  \bibfield  {author} {\bibinfo {author} {\bibfnamefont {B.}~\bibnamefont
  {Keimer}}, \bibinfo {author} {\bibfnamefont {S.~A.}\ \bibnamefont
  {Kivelson}}, \bibinfo {author} {\bibfnamefont {M.~R.}\ \bibnamefont
  {Norman}}, \bibinfo {author} {\bibfnamefont {S.}~\bibnamefont {Uchida}}, \
  and\ \bibinfo {author} {\bibfnamefont {J.}~\bibnamefont {Zaanen}},\ }\href
  {\doibase 10.1038/nature14165} {\bibfield  {journal} {\bibinfo  {journal}
  {Nature}\ }\textbf {\bibinfo {volume} {518}},\ \bibinfo {pages} {179}
  (\bibinfo {year} {2015})}\BibitemShut {NoStop}%
\bibitem [{\citenamefont {Qiao}\ \emph {et~al.}(2017)\citenamefont {Qiao},
  \citenamefont {Li}, \citenamefont {Wang}, \citenamefont {Ruan}, \citenamefont
  {Ye}, \citenamefont {Cai}, \citenamefont {Hao}, \citenamefont {Yao},
  \citenamefont {Chen}, \citenamefont {Wu}, \citenamefont {Wang},\ and\
  \citenamefont {Liu}}]{Qiao2017}%
  \BibitemOpen
  \bibfield  {author} {\bibinfo {author} {\bibfnamefont {S.}~\bibnamefont
  {Qiao}}, \bibinfo {author} {\bibfnamefont {X.}~\bibnamefont {Li}}, \bibinfo
  {author} {\bibfnamefont {N.}~\bibnamefont {Wang}}, \bibinfo {author}
  {\bibfnamefont {W.}~\bibnamefont {Ruan}}, \bibinfo {author} {\bibfnamefont
  {C.}~\bibnamefont {Ye}}, \bibinfo {author} {\bibfnamefont {P.}~\bibnamefont
  {Cai}}, \bibinfo {author} {\bibfnamefont {Z.}~\bibnamefont {Hao}}, \bibinfo
  {author} {\bibfnamefont {H.}~\bibnamefont {Yao}}, \bibinfo {author}
  {\bibfnamefont {X.}~\bibnamefont {Chen}}, \bibinfo {author} {\bibfnamefont
  {J.}~\bibnamefont {Wu}}, \bibinfo {author} {\bibfnamefont {Y.}~\bibnamefont
  {Wang}}, \ and\ \bibinfo {author} {\bibfnamefont {Z.}~\bibnamefont {Liu}},\
  }\href {\doibase 10.1103/PhysRevX.7.041054} {\bibfield  {journal} {\bibinfo
  {journal} {Physical Review X}\ }\textbf {\bibinfo {volume} {7}},\ \bibinfo
  {pages} {1} (\bibinfo {year} {2017})}\BibitemShut {NoStop}%
\bibitem [{\citenamefont {Calandra}(2018)}]{Calandra2018}%
  \BibitemOpen
  \bibfield  {author} {\bibinfo {author} {\bibfnamefont {M.}~\bibnamefont
  {Calandra}},\ }\href {\doibase 10.1103/PhysRevLett.121.026401} {\bibfield
  {journal} {\bibinfo  {journal} {Physical Review Letters}\ }\textbf {\bibinfo
  {volume} {121}},\ \bibinfo {pages} {26401} (\bibinfo {year} {2018})},\
  \Eprint {http://arxiv.org/abs/1803.08361} {arXiv:1803.08361} \BibitemShut
  {NoStop}%
\bibitem [{\citenamefont {Pasquier}\ and\ \citenamefont
  {Yazyev}(2018)}]{Pasquier2018}%
  \BibitemOpen
  \bibfield  {author} {\bibinfo {author} {\bibfnamefont {D.}~\bibnamefont
  {Pasquier}}\ and\ \bibinfo {author} {\bibfnamefont {O.~V.}\ \bibnamefont
  {Yazyev}},\ }\href {\doibase 10.1103/PhysRevB.98.045114} {\bibfield
  {journal} {\bibinfo  {journal} {Physical Review B}\ }\textbf {\bibinfo
  {volume} {98}},\ \bibinfo {pages} {1} (\bibinfo {year} {2018})},\ \Eprint
  {http://arxiv.org/abs/1803.10727} {arXiv:1803.10727} \BibitemShut {NoStop}%
\bibitem [{\citenamefont {Hubbard}(1963)}]{Hubbard1963}%
  \BibitemOpen
  \bibfield  {author} {\bibinfo {author} {\bibfnamefont {J.}~\bibnamefont
  {Hubbard}},\ }\href {\doibase 10.1098/rspa.1963.0204} {\bibfield  {journal}
  {\bibinfo  {journal} {Proceedings of the Royal Society A: Mathematical,
  Physical and Engineering Sciences}\ }\textbf {\bibinfo {volume} {276}},\
  \bibinfo {pages} {238} (\bibinfo {year} {1963})}\BibitemShut {NoStop}%
\bibitem [{\citenamefont {Spa{\L}ek}(2007)}]{Spaek2007}%
  \BibitemOpen
  \bibfield  {author} {\bibinfo {author} {\bibfnamefont {J.}~\bibnamefont
  {Spa{\L}ek}},\ }\href {\doibase 10.12693/APhysPolA.111.409} {\bibfield
  {journal} {\bibinfo  {journal} {Acta Physica Polonica A}\ }\textbf {\bibinfo
  {volume} {111}},\ \bibinfo {pages} {409} (\bibinfo {year} {2007})},\ \Eprint
  {http://arxiv.org/abs/0706.4236} {arXiv:0706.4236} \BibitemShut {NoStop}%
\bibitem [{\citenamefont {Altland}\ and\ \citenamefont
  {Simons}(2006)}]{Altland2006}%
  \BibitemOpen
  \bibfield  {author} {\bibinfo {author} {\bibfnamefont {A.}~\bibnamefont
  {Altland}}\ and\ \bibinfo {author} {\bibfnamefont {B.}~\bibnamefont
  {Simons}},\ }\href {\doibase 10.1007/BFb0120123} {\emph {\bibinfo {title}
  {Book}}},\ Vol.\ \bibinfo {volume} {115}\ (\bibinfo {year} {2006})\ p.\
  \bibinfo {pages} {786}\BibitemShut {NoStop}%
\bibitem [{\citenamefont {Bardeen}\ \emph {et~al.}(1957)\citenamefont
  {Bardeen}, \citenamefont {Cooper},\ and\ \citenamefont
  {Schrieffer}}]{Bardeen1957}%
  \BibitemOpen
  \bibfield  {author} {\bibinfo {author} {\bibfnamefont {J.}~\bibnamefont
  {Bardeen}}, \bibinfo {author} {\bibfnamefont {L.~N.}\ \bibnamefont {Cooper}},
  \ and\ \bibinfo {author} {\bibfnamefont {J.~R.}\ \bibnamefont {Schrieffer}},\
  }\href {\doibase 10.1103/PhysRev.108.1175} {\bibfield  {journal} {\bibinfo
  {journal} {Physical Review}\ }\textbf {\bibinfo {volume} {108}},\ \bibinfo
  {pages} {1175} (\bibinfo {year} {1957})}\BibitemShut {NoStop}%
\bibitem [{\citenamefont {Donos}\ and\ \citenamefont
  {Hartnoll}(2012)}]{Donos2012}%
  \BibitemOpen
  \bibfield  {author} {\bibinfo {author} {\bibfnamefont {A.}~\bibnamefont
  {Donos}}\ and\ \bibinfo {author} {\bibfnamefont {S.~A.}\ \bibnamefont
  {Hartnoll}},\ }\href {http://arxiv.org/abs/1212.2998} {\bibfield  {journal}
  {\bibinfo  {journal} {arXiv}\ ,\ \bibinfo {pages} {5}} (\bibinfo {year}
  {2012})},\ \Eprint {http://arxiv.org/abs/1212.2998} {arXiv:1212.2998}
  \BibitemShut {NoStop}%
\bibitem [{\citenamefont {Parke}\ and\ \citenamefont
  {Taylor}(1986)}]{Parke1986}%
  \BibitemOpen
  \bibfield  {author} {\bibinfo {author} {\bibfnamefont {S.~J.}\ \bibnamefont
  {Parke}}\ and\ \bibinfo {author} {\bibfnamefont {T.~R.}\ \bibnamefont
  {Taylor}},\ }\href {\doibase 10.1103/PhysRevLett.56.2459} {\bibfield
  {journal} {\bibinfo  {journal} {Physical Review Letters}\ }\textbf {\bibinfo
  {volume} {56}},\ \bibinfo {pages} {2459} (\bibinfo {year}
  {1986})}\BibitemShut {NoStop}%
\bibitem [{\citenamefont {Dixon}(2013)}]{Dixon2013}%
  \BibitemOpen
  \bibfield  {author} {\bibinfo {author} {\bibfnamefont {L.~J.}\ \bibnamefont
  {Dixon}},\ }\href {\doibase 10.5170/CERN-2014-008.31} {\  (\bibinfo {year}
  {2013}),\ 10.5170/CERN-2014-008.31},\ \Eprint
  {http://arxiv.org/abs/1310.5353} {arXiv:1310.5353} \BibitemShut {NoStop}%
\bibitem [{\citenamefont {Arkani-Hamed}\ \emph {et~al.}(2017)\citenamefont
  {Arkani-Hamed}, \citenamefont {Huang},\ and\ \citenamefont
  {Huang}}]{Arkani-Hamed2017}%
  \BibitemOpen
  \bibfield  {author} {\bibinfo {author} {\bibfnamefont {N.}~\bibnamefont
  {Arkani-Hamed}}, \bibinfo {author} {\bibfnamefont {T.-C.}\ \bibnamefont
  {Huang}}, \ and\ \bibinfo {author} {\bibfnamefont {Y.-t.}\ \bibnamefont
  {Huang}},\ }\href {http://arxiv.org/abs/1709.04891} {\  (\bibinfo {year}
  {2017})},\ \Eprint {http://arxiv.org/abs/1709.04891} {arXiv:1709.04891}
  \BibitemShut {NoStop}%
\bibitem [{\citenamefont {Britto}\ \emph {et~al.}(2005)\citenamefont {Britto},
  \citenamefont {Cachazo},\ and\ \citenamefont {Feng}}]{Britto2005}%
  \BibitemOpen
  \bibfield  {author} {\bibinfo {author} {\bibfnamefont {R.}~\bibnamefont
  {Britto}}, \bibinfo {author} {\bibfnamefont {F.}~\bibnamefont {Cachazo}}, \
  and\ \bibinfo {author} {\bibfnamefont {B.}~\bibnamefont {Feng}},\ }\href
  {\doibase 10.1016/j.nuclphysb.2005.02.030} {\bibfield  {journal} {\bibinfo
  {journal} {Nuclear Physics B}\ }\textbf {\bibinfo {volume} {715}},\ \bibinfo
  {pages} {499} (\bibinfo {year} {2005})},\ \Eprint
  {http://arxiv.org/abs/0412308} {arXiv:0412308 [hep-th]} \BibitemShut
  {NoStop}%
\bibitem [{\citenamefont {Morin}(1959)}]{Morin1959}%
  \BibitemOpen
  \bibfield  {author} {\bibinfo {author} {\bibfnamefont {F.~J.}\ \bibnamefont
  {Morin}},\ }\href@noop {} {\bibfield  {journal} {\bibinfo  {journal} {Phys.
  Rev. Lett.}\ }\textbf {\bibinfo {volume} {3}},\ \bibinfo {pages} {2}
  (\bibinfo {year} {1959})}\BibitemShut {NoStop}%
\bibitem [{\citenamefont {Eyert}(2002)}]{Eyert2002}%
  \BibitemOpen
  \bibfield  {author} {\bibinfo {author} {\bibfnamefont {V.}~\bibnamefont
  {Eyert}},\ }\href@noop {} {\bibfield  {journal} {\bibinfo  {journal} {Annalen
  der Physik}\ }\textbf {\bibinfo {volume} {11}},\ \bibinfo {pages} {650}
  (\bibinfo {year} {2002})}\BibitemShut {NoStop}%
\bibitem [{\citenamefont {Marezio}\ \emph {et~al.}(1971)\citenamefont
  {Marezio}, \citenamefont {McWhan}, \citenamefont {Remeika},\ and\
  \citenamefont {Dernier}}]{Marezio1971}%
  \BibitemOpen
  \bibfield  {author} {\bibinfo {author} {\bibfnamefont {M.}~\bibnamefont
  {Marezio}}, \bibinfo {author} {\bibfnamefont {D.~B.}\ \bibnamefont {McWhan}},
  \bibinfo {author} {\bibfnamefont {J.~P.}\ \bibnamefont {Remeika}}, \ and\
  \bibinfo {author} {\bibfnamefont {P.~D.}\ \bibnamefont {Dernier}},\
  }\href@noop {} {\bibfield  {journal} {\bibinfo  {journal} {Phys Rev B}\
  }\textbf {\bibinfo {volume} {91}},\ \bibinfo {pages} {2541} (\bibinfo {year}
  {1971})}\BibitemShut {NoStop}%
\bibitem [{\citenamefont {Booth}\ and\ \citenamefont
  {Casey}(2009)}]{Booth2009}%
  \BibitemOpen
  \bibfield  {author} {\bibinfo {author} {\bibfnamefont {J.~M.}\ \bibnamefont
  {Booth}}\ and\ \bibinfo {author} {\bibfnamefont {P.~S.}\ \bibnamefont
  {Casey}},\ }\href {\doibase 10.1021/am900322b} {\bibfield  {journal}
  {\bibinfo  {journal} {ACS Appl. Mater. Interfaces}\ }\textbf {\bibinfo
  {volume} {1}},\ \bibinfo {pages} {1899} (\bibinfo {year} {2009})}\BibitemShut
  {NoStop}%
\bibitem [{\citenamefont {Pouget}\ and\ \citenamefont
  {Launois}(1976)}]{Pouget1976}%
  \BibitemOpen
  \bibfield  {author} {\bibinfo {author} {\bibfnamefont {J.~P.}\ \bibnamefont
  {Pouget}}\ and\ \bibinfo {author} {\bibfnamefont {H.}~\bibnamefont
  {Launois}},\ }\href@noop {} {\bibfield  {journal} {\bibinfo  {journal}
  {Journal de Physique}\ }\textbf {\bibinfo {volume} {C4}},\ \bibinfo {pages}
  {49} (\bibinfo {year} {1976})}\BibitemShut {NoStop}%
\bibitem [{\citenamefont {Budai}\ \emph {et~al.}(2014)\citenamefont {Budai},
  \citenamefont {Hong}, \citenamefont {Manley}, \citenamefont {Specht},
  \citenamefont {Li}, \citenamefont {Tischler}, \citenamefont {Abernathy},
  \citenamefont {Said}, \citenamefont {Leu}, \citenamefont {Boatner},
  \citenamefont {Mcqueeney},\ and\ \citenamefont {Delaire}}]{Budai2014}%
  \BibitemOpen
  \bibfield  {author} {\bibinfo {author} {\bibfnamefont {J.~D.}\ \bibnamefont
  {Budai}}, \bibinfo {author} {\bibfnamefont {J.}~\bibnamefont {Hong}},
  \bibinfo {author} {\bibfnamefont {M.~E.}\ \bibnamefont {Manley}}, \bibinfo
  {author} {\bibfnamefont {E.~D.}\ \bibnamefont {Specht}}, \bibinfo {author}
  {\bibfnamefont {C.~W.}\ \bibnamefont {Li}}, \bibinfo {author} {\bibfnamefont
  {J.~Z.}\ \bibnamefont {Tischler}}, \bibinfo {author} {\bibfnamefont {D.~L.}\
  \bibnamefont {Abernathy}}, \bibinfo {author} {\bibfnamefont {A.~H.}\
  \bibnamefont {Said}}, \bibinfo {author} {\bibfnamefont {B.~M.}\ \bibnamefont
  {Leu}}, \bibinfo {author} {\bibfnamefont {L.~a.}\ \bibnamefont {Boatner}},
  \bibinfo {author} {\bibfnamefont {R.~J.}\ \bibnamefont {Mcqueeney}}, \ and\
  \bibinfo {author} {\bibfnamefont {O.}~\bibnamefont {Delaire}},\ }\href
  {\doibase 10.1038/nature13865} {\bibfield  {journal} {\bibinfo  {journal}
  {Nature}\ }\textbf {\bibinfo {volume} {515}},\ \bibinfo {pages} {535}
  (\bibinfo {year} {2014})}\BibitemShut {NoStop}%
\bibitem [{\citenamefont {Aschroft}\ and\ \citenamefont
  {Mermin}(2011)}]{Ashcroft1976}%
  \BibitemOpen
  \bibfield  {author} {\bibinfo {author} {\bibfnamefont {N.~W.}\ \bibnamefont
  {Aschroft}}\ and\ \bibinfo {author} {\bibfnamefont {N.~D.}\ \bibnamefont
  {Mermin}},\ }\href@noop {} {\emph {\bibinfo {title} {{Solid State
  Physics}}}}\ (\bibinfo  {publisher} {Cengage Learning},\ \bibinfo {year}
  {2011})\ p.\ \bibinfo {pages} {439}\BibitemShut {NoStop}%
\bibitem [{\citenamefont {Peskin}\ and\ \citenamefont
  {Schroeder}(2016)}]{Peskin2016}%
  \BibitemOpen
  \bibfield  {author} {\bibinfo {author} {\bibfnamefont {M.~E.}\ \bibnamefont
  {Peskin}}\ and\ \bibinfo {author} {\bibfnamefont {D.~V.}\ \bibnamefont
  {Schroeder}},\ }\href@noop {} {\emph {\bibinfo {title} {{An Introduction to
  Quantum Field Theory}}}}\ (\bibinfo  {publisher} {Westview Press},\ \bibinfo
  {year} {2016})\ pp.\ \bibinfo {pages} {238--244}\BibitemShut {NoStop}%
\bibitem [{\citenamefont {Shishkin}\ and\ \citenamefont
  {Kresse}(2006)}]{Shishkin2006}%
  \BibitemOpen
  \bibfield  {author} {\bibinfo {author} {\bibfnamefont {M.}~\bibnamefont
  {Shishkin}}\ and\ \bibinfo {author} {\bibfnamefont {G.}~\bibnamefont
  {Kresse}},\ }\href {\doibase 10.1103/PhysRevB.74.035101} {\bibfield
  {journal} {\bibinfo  {journal} {Phys. Rev. B}\ }\textbf {\bibinfo {volume}
  {74}},\ \bibinfo {pages} {35101} (\bibinfo {year} {2006})}\BibitemShut
  {NoStop}%
\bibitem [{\citenamefont {Shishkin}\ and\ \citenamefont
  {Kresse}(2007)}]{Shishkin2007}%
  \BibitemOpen
  \bibfield  {author} {\bibinfo {author} {\bibfnamefont {M.}~\bibnamefont
  {Shishkin}}\ and\ \bibinfo {author} {\bibfnamefont {G.}~\bibnamefont
  {Kresse}},\ }\href {\doibase 10.1103/PhysRevB.75.235102} {\bibfield
  {journal} {\bibinfo  {journal} {Phys. Rev. B}\ }\textbf {\bibinfo {volume}
  {75}},\ \bibinfo {pages} {235102} (\bibinfo {year} {2007})}\BibitemShut
  {NoStop}%
\bibitem [{\citenamefont {Kresse}\ and\ \citenamefont
  {Furthm{\"{u}}ller}(1996)}]{Kresse1996}%
  \BibitemOpen
  \bibfield  {author} {\bibinfo {author} {\bibfnamefont {G.}~\bibnamefont
  {Kresse}}\ and\ \bibinfo {author} {\bibfnamefont {J.}~\bibnamefont
  {Furthm{\"{u}}ller}},\ }\href@noop {} {\bibfield  {journal} {\bibinfo
  {journal} {Phys. Rev. B}\ }\textbf {\bibinfo {volume} {54}},\ \bibinfo
  {pages} {11169} (\bibinfo {year} {1996})}\BibitemShut {NoStop}%
\bibitem [{\citenamefont {Kohn}\ and\ \citenamefont {Sham}(1965)}]{Kohn1965}%
  \BibitemOpen
  \bibfield  {author} {\bibinfo {author} {\bibfnamefont {W.}~\bibnamefont
  {Kohn}}\ and\ \bibinfo {author} {\bibfnamefont {L.~J.}\ \bibnamefont
  {Sham}},\ }\href@noop {} {\bibfield  {journal} {\bibinfo  {journal} {Phys.
  Rev.}\ }\textbf {\bibinfo {volume} {140}},\ \bibinfo {pages} {1133} (\bibinfo
  {year} {1965})}\BibitemShut {NoStop}%
\bibitem [{\citenamefont {Perdew}\ \emph {et~al.}(1996)\citenamefont {Perdew},
  \citenamefont {Burke},\ and\ \citenamefont {Ernzerhof}}]{Perdew1996}%
  \BibitemOpen
  \bibfield  {author} {\bibinfo {author} {\bibfnamefont {J.~P.}\ \bibnamefont
  {Perdew}}, \bibinfo {author} {\bibfnamefont {K.}~\bibnamefont {Burke}}, \
  and\ \bibinfo {author} {\bibfnamefont {M.}~\bibnamefont {Ernzerhof}},\
  }\href@noop {} {\bibfield  {journal} {\bibinfo  {journal} {Phys. Rev. Lett.}\
  }\textbf {\bibinfo {volume} {77}},\ \bibinfo {pages} {3865} (\bibinfo {year}
  {1996})}\BibitemShut {NoStop}%
\bibitem [{\citenamefont {Monkhorst}\ and\ \citenamefont
  {Pack}(1976)}]{Monkhorst1976}%
  \BibitemOpen
  \bibfield  {author} {\bibinfo {author} {\bibfnamefont {H.~J.}\ \bibnamefont
  {Monkhorst}}\ and\ \bibinfo {author} {\bibfnamefont {J.~D.}\ \bibnamefont
  {Pack}},\ }\href@noop {} {\bibfield  {journal} {\bibinfo  {journal} {Phys.
  Rev. B}\ }\textbf {\bibinfo {volume} {13}},\ \bibinfo {pages} {5188}
  (\bibinfo {year} {1976})}\BibitemShut {NoStop}%
\bibitem [{\citenamefont {Blochl}\ \emph {et~al.}(1994)\citenamefont {Blochl},
  \citenamefont {Jepsen},\ and\ \citenamefont {Andersen}}]{Bloechl1994}%
  \BibitemOpen
  \bibfield  {author} {\bibinfo {author} {\bibfnamefont {P.~E.}\ \bibnamefont
  {Blochl}}, \bibinfo {author} {\bibfnamefont {O.}~\bibnamefont {Jepsen}}, \
  and\ \bibinfo {author} {\bibfnamefont {O.~K.}\ \bibnamefont {Andersen}},\
  }\href@noop {} {\bibfield  {journal} {\bibinfo  {journal} {Phys. Rev. B}\
  }\textbf {\bibinfo {volume} {49}},\ \bibinfo {pages} {16223} (\bibinfo {year}
  {1994})}\BibitemShut {NoStop}%
\bibitem [{\citenamefont {Longo}\ and\ \citenamefont
  {Kierkegaard}(1970)}]{Longo1970}%
  \BibitemOpen
  \bibfield  {author} {\bibinfo {author} {\bibfnamefont {J.~M.}\ \bibnamefont
  {Longo}}\ and\ \bibinfo {author} {\bibfnamefont {P.}~\bibnamefont
  {Kierkegaard}},\ }\href@noop {} {\bibfield  {journal} {\bibinfo  {journal}
  {Acta Chemica Scandinavica}\ }\textbf {\bibinfo {volume} {24}},\ \bibinfo
  {pages} {420} (\bibinfo {year} {1970})}\BibitemShut {NoStop}%
\bibitem [{\citenamefont {Methfessel}\ and\ \citenamefont
  {Paxton}(1989)}]{Methfessel1989}%
  \BibitemOpen
  \bibfield  {author} {\bibinfo {author} {\bibfnamefont {M.}~\bibnamefont
  {Methfessel}}\ and\ \bibinfo {author} {\bibfnamefont {A.~T.}\ \bibnamefont
  {Paxton}},\ }\href@noop {} {\bibfield  {journal} {\bibinfo  {journal}
  {Physical review. B}\ }\textbf {\bibinfo {volume} {40}},\ \bibinfo {pages}
  {3616} (\bibinfo {year} {1989})}\BibitemShut {NoStop}%
\bibitem [{\citenamefont {Gonze}(1995)}]{Gonze1995b}%
  \BibitemOpen
  \bibfield  {author} {\bibinfo {author} {\bibfnamefont {X.}~\bibnamefont
  {Gonze}},\ }\href {\doibase 10.1103/PhysRevA.52.1096} {\bibfield  {journal}
  {\bibinfo  {journal} {Phys. Rev. A}\ }\textbf {\bibinfo {volume} {52}},\
  \bibinfo {pages} {1096} (\bibinfo {year} {1995})}\BibitemShut {NoStop}%
\bibitem [{\citenamefont {Togo}\ and\ \citenamefont {Tanaka}(2015)}]{Togo2015}%
  \BibitemOpen
  \bibfield  {author} {\bibinfo {author} {\bibfnamefont {A.}~\bibnamefont
  {Togo}}\ and\ \bibinfo {author} {\bibfnamefont {I.}~\bibnamefont {Tanaka}},\
  }\href@noop {} {\bibfield  {journal} {\bibinfo  {journal} {Scr. Mater.}\
  }\textbf {\bibinfo {volume} {108}},\ \bibinfo {pages} {1} (\bibinfo {year}
  {2015})}\BibitemShut {NoStop}%
\end{thebibliography}%
\end{document}